\documentclass[journal]{IEEEtran}
\usepackage{balance}
\usepackage{graphicx}
\usepackage{bm}
\usepackage{amsmath}
\usepackage{amsfonts}
\usepackage{amsthm}
\usepackage{amssymb}
\usepackage{color}
\usepackage{mathrsfs}
\usepackage{upgreek}
\usepackage[utf8]{inputenc}
\usepackage{tikz}
\usetikzlibrary{arrows, calc, decorations.markings, positioning}
\usepackage{cite}
\usepackage{booktabs}
\usepackage{multirow}
\usepackage{amsmath}
\usepackage{amsthm}
\usepackage{algorithm}
\usepackage{algorithmic}
\usepackage{hyperref}
\usepackage{flushend}
\usepackage{midfloat}
\usepackage{tabularx}
\usepackage{threeparttable}
\ifCLASSOPTIONcompsoc
 \usepackage[caption=false,font=normalsize,labelfont=sf,textfont=sf]{subfig}
\else
 \usepackage[caption=false,font=footnotesize]{subfig}
\fi

\usepackage{hyperref}					
\hypersetup{colorlinks,
	linkcolor=blue,%
	anchorcolor=blue,
    citecolor=blue}

\begin{document}
\title{{Sensing With Communication Signals:\\ From Information Theory to Signal Processing}
}
\author{
	{
	Fan Liu,~\IEEEmembership{Senior Member,~IEEE}, Ya-Feng Liu,~\IEEEmembership{Senior Member,~IEEE}, Yuanhao Cui,~\IEEEmembership{Member, IEEE}, \\Christos Masouros,~\IEEEmembership{Fellow,~IEEE}, Jie Xu,~\IEEEmembership{Fellow, IEEE}, Tony Xiao Han,~\IEEEmembership{Senior Member, IEEE}, \\Stefano Buzzi,~\IEEEmembership{Senior Member, IEEE}, Yonina C. Eldar,~\IEEEmembership{Fellow, IEEE}, and Shi Jin,~\IEEEmembership{Fellow, IEEE}
\thanks{F. Liu and S. Jin are with the National Mobile Communications Research Laboratory, School of Information Science and Engineering, Southeast University, Nanjing 210096, China (e-mail: fan.liu@seu.edu.cn, jinshi@seu.edu.cn).}
\thanks{Y.-F. Liu is with the State Key Laboratory of Scientific and Engineering Computing, Institute of Computational Mathematicsand Scientific/Engineering Computing, Academy of Mathematics and Systems Science, Chinese Academy of Sciences, Beijing 100190, China (e-mail: yafliu@lsec.cc.ac.cn).}
\thanks{Y. Cui is with the Department of Communication Engineering, Beijing University of Posts and Telecommunications, Beijing 100876, China (e-mail: cuiyuanhao@bupt.edu.cn).}
\thanks{C. Masouros is with the Department of Electronic and Electrical Engineering, University College London, London, WC1E 7JE, UK (e-mail: chris.masouros@ieee.org).}
\thanks{J. Xu is with the School of Science and Engineering, the
Shenzhen Future Network of Intelligence Institute (FNii-Shenzhen), and the
Guangdong Provincial Key Laboratory of Future Networks of Intelligence,
The Chinese University of Hong Kong (Shenzhen), Guangdong 518172,
China (e-mail: xujie@cuhk.edu.cn).}
\thanks{T. X. Han is with Huawei Technologies Co., Ltd (email: tony.hanxiao@huawei.com).}
\thanks{S. Buzzi is with the Department of Electrical and Information Engineering, University of Cassino and Southern Lazio, I-03043 Cassino, Italy, with the Department of Electronics, Information and
Bioengineering, Politecnico di Milano, I-20133 Milano, Italy, and with the Consorzio Nazionale Interuniversitario per le Telecomunicazioni (CNIT), I-43124 Parma, Italy (e-mail: buzzi@unicas.it).}
\thanks{Y. C. Eldar is with the Faculty of Mathematics and Computer Science, Weizmann Institute of Science, Rehovot, Israel (e-mail: yonina.eldar@weizmann.ac.il).}
} 

}
\maketitle

\begin{abstract}
The Integrated Sensing and Communications (ISAC) paradigm is anticipated to be a cornerstone of the upcoming 6G networks. In order to optimize the use of wireless resources, 6G ISAC systems need to harness the communication data payload signals, which are inherently random, for both sensing and communication (S\&C) purposes. This tutorial paper provides a comprehensive technical overview of the fundamental theory and signal processing methodologies for ISAC transmission with random communication signals. We begin by introducing the deterministic-random tradeoff (DRT) between S\&C from an information-theoretic perspective, emphasizing the need for specialized signal processing techniques tailored to random ISAC signals. Building on this foundation, we review the core signal models and processing pipelines for communication-centric ISAC systems, and analyze the average squared auto-correlation function (ACF) of random ISAC signals, which serves as a fundamental performance metric for multi-target ranging tasks. Drawing insights from these theoretical results, we outline the design principles for the three key components of communication-centric ISAC systems: modulation schemes, constellation design, and pulse shaping filters. The goal is to either enhance sensing performance without compromising communication efficiency or to establish a scalable tradeoff between the two. We then extend our analysis from a single-antenna ISAC system to its multi-antenna counterpart, discussing recent advancements in multi-input multi-output (MIMO) precoding techniques specifically designed for random ISAC signals. We conclude by highlighting several open challenges and future research directions in the field of sensing with communication signals.
\end{abstract}

\section{Introduction}\label{intro_sec}
\subsection{Background and Motivation}
\IEEEPARstart{N}{ext}-generation wireless networks (5G-Advanced (5G-A) and 6G) are increasingly recognized as a pivotal enabler for a broad spectrum of emerging applications, including the Digital Twin, Metaverse, Smart Cities, Industrial Internet-of-Things (IoT), and the Low-Altitude Economy powered by Unmanned Aerial Vehicles (UAVs)  \cite{Chafii2023CST,saad2019vision}. These applications promise to revolutionize industries and societies by providing advanced digital experiences, real-time monitoring, and intelligent automation. In May 2023, the International Telecommunications Union (ITU) successfully completed the Recommendation Framework for IMT-2030, which is commonly referred to as the {\textit{global 6G vision}}  \cite{ITU2023}. This achievement marks the formal initiation of the 6G standardization process, laying the foundation for the development of next-generation communication systems. Among the six key usage scenarios identified by the ITU, Integrated Sensing and Communications (ISAC) stands out as a particularly transformative innovation \cite{cui2021integrating,9585321}. ISAC is envisioned to offer integrated solutions that combine wireless sensing and communication (S\&C) functions in a seamless and efficient manner, thereby enhancing the performance and capabilities of 6G networks  \cite{9737357,9585321}. By incorporating native sensing functionalities directly into the communication infrastructure, ISAC-enabled cellular networks will further unlock distributed sensing of unprecedented scale. In doing so, it will support another one of key 6G usage scenarios, that of integrated artificial intelligence (AI) and communications, by generating the necessary volume of sensory data to build the networked intelligence.

ISAC technologies can be conceptualized in a number of progressive stages \cite{10188491}. The initial stage focuses on the independent use of spectral resources by S\&C systems, ensuring no interference between the two. In the subsequent stage, both S\&C functions are consolidated onto a shared RF front-end. In the third stage, namely, fully integrated ISAC systems, S\&C functions are performed on a unified hardware platform using a shared waveform within the same frequency band \cite{9737357}. In such a system, a single radio signal is transmitted to both deliver data information to communication users, as well as to acquire critical information from the returned echoes, e.g., range, angle, velocity, trajectory, size and shape of targets of interest, or even image of the surrounding environment. This integration poses considerable challenges at the physical layer (PHY), where innovative signaling schemes are crucial for supporting higher-layer ISAC applications. Among various design strategies, three primary approaches have garnered significant attention from both academia and industry, as outlined below.
\subsubsection{Sensing-Centric Design}
Sensing-centric design focuses on incorporating communication bits into legacy radar waveforms, which is often referred to as the {\textit{information embedding}} method \cite{hassanien2016signaling,8828023,7347464,10018010}. Taking chirp waveforms as an example, this can be achieved by representing communication symbols through the amplitude, phase, frequency, or even the chirp rate of the waveform \cite{roberton2003integrated,saddik2007ultra,10018010}. Additionally, for a MIMO radar system, useful information can also be embedded in the spatial domain, through techniques like sidelobe control of the beampattern or index modulation \cite{Ma2021FRaC,9093221}, where waveforms with different carrier frequencies are shuffled across multiple antennas \cite{7347464,7485066}. However, since most sensing-centric methods rely on slow-time coding to avoid disrupting the structure of radar waveforms for preserving the ISAC system's sensing capability, these approaches typically result in lower data rates, constrained by the pulse repetition frequency (PRF) \cite{zheng2019radar}.
\subsubsection{Communication-Centric Design}
In contrast to the sensing-centric design, the communication-centric approach directly implements the sensing functionality within existing communication signaling formats. Some of the early efforts in this area can be traced back to the code-division multiple access (CDMA) era, where Oppermann sequences were used to achieve both communication and radar functionalities \cite{4753277}. A more notable example is the orthogonal frequency division multiplexing (OFDM) waveform \cite{9727202,10298608}, whose sensing capabilities were demonstrated in the seminal paper \cite{sturm2011waveform}. As a further step, orthogonal time-frequency space (OTFS) modulation has been considered as a potential candidate waveform for 6G ISAC applications due to its operation in the delay-Doppler domain, which directly corresponds to the key parameters of radar targets \cite{9109735,10638525,10463758}. In recent developments, a novel communication waveform known as affine frequency division multiplexing (AFDM) modulation was proposed to enhance both S\&C performance, especially in high-mobility scenarios \cite{10087310,10439996}. While in principle, any communication signals can be leveraged for ISAC, maintaining the communication functionality with a minimal impact, it may lead to degraded and difficult-to-tune sensing performance, as these signals are not specifically designed for target detection or estimation.
\subsubsection{Joint Design}
Aiming to strike a scalable tradeoff between S\&C, the joint design methodology develops ISAC waveforms from the ground up, rather than relying solely on conventional radar or communication signals \cite{9127852,9345999}. In this framework, the ISAC signaling problem is formulated as a multi-objective optimization, where specific S\&C performance metrics are incorporated as cost functions or constraints. This may involve minimizing the inter-user interference for communication, subject to MIMO radar beampattern constraints \cite{liuTSP2018,8288677,9724205,9124713}, or minimizing the Cram\'er-Rao Bound (CRB) for sensing, subject to per-user signal-to-interference-plus-noise ratio (SINR), or transmission rate requirements for communication \cite{9652071,10251151,10217169}. More recently, various techniques have been introduced for ISAC waveform design to address issues such as hardware non-ideality \cite{10494366,10770016,10839033,10845891}. In general, the joint design approach offers significant flexibility for ISAC signaling, but comes at the cost of considerably higher complexity compared to the other two designs.

Among the three design philosophies, the communication-centric ISAC design holds more promise for practical deployment in 5G-A and 6G networks, primarily due to its low implementation costs and full compatibility with existing cellular infrastructure \cite{10012421}. Indeed, ISAC standardization is progressing well within the 3rd Generation Partnership Project (3GPP), which builds upon the 5G New Radio (5G NR) protocols. For instance, a key focus of 3GPP Release 18 (Rel-18) is on improving device positioning \cite{Cha_Rel18}, crucial for ISAC and wireless sensing in 5G-A. As a further step, a Technical Report TR 22.837 was introduced towards Rel-19 in April 2022, identifying 32 ISAC use cases \cite{TR22.837}. In August 2023, the Technical Specification TS 22.137 outlined the service requirements for wireless sensing, detailing eight key performance indicators (KPIs) \cite{TS22.137}. Additionally, in December 2023, a study on channel modeling for ISAC was approved \cite{3gppISAC_channel_model}. Complementary work is also underway in the European Telecommunications Standards Institute (ETSI), focusing on ISAC-specific use cases and security. Meanwhile, IEEE 802.11bf aims to enhance WLAN standards for sensing \cite{10547188}. On top of that, air interface technologies for ISAC are expected to be studied in 3GPP Rel-20 starting in 2025 onward, and will be finalized in the first set of 3GPP 6G technical specifications under Rel-21 \cite{Lin_Rel19}. As part of this evolution, research on PHY signaling and processing techniques for communication-centric ISAC will become increasingly crucial.

\subsection{Sensing With Random Communication Signals}
The current 5G NR ISAC signaling framework relies heavily on reference signals, known as ``pilots'', embedded within the NR frame structure. For instance, as defined in 3GPP Rel-18, Reduced-Capability (RedCap) devices use positioning reference signals (PRS) for performing downlink positioning measurements, while transmitting sounding reference signals (SRS) for uplink positioning \cite{Cha_Rel18,9921271}. In addition to PRS and SRS, other PHY reference signals have also been explored for sensing purposes, such as channel state information reference signals (CSI-RS), demodulation reference signals (DMRS), and synchronization signals \cite{10561589,10457036,9746355}. These signals are typically generated from pseudo-random sequences, including Zadoff-Chu sequences, $m$-sequences, and Gold sequences, which possess beneficial auto- and cross-correlation properties \cite{9746355}. Despite their advantages, PHY reference signals occupy only up to 10\%-15\% of time-frequency resources, which severely limits their ability to provide superior sensing performance required in emerging ISAC use cases. Scaling up the sensing performance would inevitably involve collecting sensing samples over multiple frames, introducing excessive complexity and latency. To fill in this gap, a promising approach is to re-purpose data payloads, which account for over 85\% of the available resources, for both S\&C tasks. This approach significantly enhances the range-Doppler resolution as well as target detection and estimation performance by fully exploiting the entire bandwidth and time duration of communication signals.

While leveraging data payload signals for ISAC offers significant performance gains over conventional pilot-only schemes, these signals are not inherently designed for sensing applications, leading to critical challenges in implementing communication-centric ISAC systems. First, unlike pseudo-random sequences discussed earlier, data payload signals are \textit{random signals} that carry useful information \cite{cover1999elements}. These signals are generated randomly from specific codebooks, with their structure determined by the distribution of information sources. Recent advancements in ISAC information theory have underscored a key distinction between the requirements for S\&C. That is, communication systems rely on random signals to efficiently convey information, while radar sensing systems demand deterministic signals with favorable ambiguity properties. Consequently, the randomness embedded in ISAC signals improves the communication rate but deteriorates sensing performance, giving rise to the deterministic-random tradeoff (DRT) between S\&C \cite{10206462,Xiong_TIT,10471902,9785593}. This tradeoff, particularly concerning the input distribution of ISAC signals \cite{10744042}, poses a significant challenge in characterzing the Pareto performance boundary for S\&C. In this context, the work in \cite{Xiong_TIT} examined a basic point-to-point (P2P) ISAC system operating over vector Gaussian channels, and evaluated the achievable S\&C performance at communication- and sensing-optimal operating points, respectively, providing valuable theoretical insights into the design of more sophisticated ISAC systems by leveraging the DRT. 

Second, beyond the foundational nature of the theoretical findings derived from ISAC information theory \cite{10206462,Xiong_TIT,10471902,9785593}, it is of practical importance to further evaluate and optimize the achievable sensing performance under real-world communication signals. Conceived primarily for data delivery, communication signals have a format fundamentally different from conventional radar signals. At its most basic level, a practical communication signal can be decomposed into the following key components \cite{proakis2008digital}:
\begin{itemize}
    \item \textbf{Channel Codes} that encode information sources into coded bit sequences to improve transmission reliability.
    \item \textbf{Constellation Symbols} mapped from bit sequences that carry information.
    \item \textbf{Orthonormal Modulation Basis} that conveys these symbols by formulating discrete-time signals.
    \item \textbf{Pulse Shaping Filter} that converts discrete time-domain samples into continuous-time signals.
    \item \textbf{MIMO Precoder} required in multi-antenna systems to enable multi-stream or multi-layer transmission.
\end{itemize}
Each of these components significantly influences the resulting sensing performance, yet their impacts remain largely unexplored. To effectively guide the development of communication-centric ISAC systems for future 6G networks, a deeper understanding of how these core elements of communication systems influence the sensing performance is essential. This understanding serves as the basic motivation of the study in this tutorial paper.

\subsection{Organization of This Paper}
In this tutorial paper, we present a comprehensive technical overview of recent advances in the fundamental theory and signal processing methodologies for P2P ISAC systems that leverage random data payload signals \cite{10206462,Xiong_TIT,10471902,9785593,liu2025iceberg,liu2024OFDM,10685511,Liao_Pulse_Shaping,10596930}. We begin by introducing the DRT between S\&C from an information-theoretic standpoint in  Sec. \ref{drt_sec}, underscoring the importance of developing tailored signal processing approaches specifically for random ISAC signals \cite{10206462,Xiong_TIT,10471902,8437621,8849242,9457571,9785593}. In particular, we review the capacity-distortion (C-D) theory for state-dependent memoryless ISAC channels \cite{9785593}, depicting the impact of input distribution for both S\&C. We then generalize our analysis to vector Gaussian channels by examining the CRB-rate tradeoff for ISAC \cite{Xiong_TIT}, which reveals the optimal distribution and structure of ISAC signal matrix at sensing- and communication-optimal points, and characterizes the achievable S\&C performance, respectively. 

Expanding on this theoretical foundation, we proceed to review the core signal models and processing pipelines for communication-centric ISAC systems in Sec. \ref{comm_centric_model_sec}. A key focus is on the auto-correlation function (ACF) of random ISAC signals, which serves as a critical metric for assessing the sensing performance in multi-target ranging applications \cite{liu2025iceberg}. Due to the inherent randomness of data payloads, analyzing the statistical property of ACF becomes essential, instead of relying on specific instances. Recent research has focused on deriving a closed-form expression for the expected squared ACF \cite{liu2025iceberg}, taking into account arbitrary modulation techniques and constellation mappings within the Nyquist pulse shaping framework. This expression is metaphorically described as an ``iceberg-in-the-sea'' structure, where the ``iceberg'' represents the squared mean of the ACF of random ISAC signals, determined by the pulse shaping filter, and the ``sea level'' corresponds to the variance, which reflects the variability introduced by the data randomness. 

Drawing insights from these results, we further overview the design principles for the three key components of communication-centric ISAC systems in Sec. \ref{waveform_design_sec}, including modulation schemes \cite{liu2024OFDM}, constellation designs \cite{10685511}, and pulse shaping filters \cite{liu2025iceberg,Liao_Pulse_Shaping}. The objective is either to improve sensing performance without sacrificing communication efficiency, or, alternatively, to establish a scalable tradeoff between the two. This balance is crucial for enabling the seamless integration of S\&C functionalities within the same system. Notably, we show that among all orthogonal linear modulation schemes, OFDM attains the lowest average ranging sidelobe level for independently and identically distributed (i.i.d.) QAM/PSK symbols \cite{liu2024OFDM}. We then review a probabilistic constellation shaping method to maximize the communication rate while further reducing the sidelobe level for OFDM signaling \cite{10685511}, followed by a Nyquist pulse design approach to reshape the ACF of ISAC signals \cite{liu2025iceberg,Liao_Pulse_Shaping}.
Furthermore, we extend the discussion to more complex MIMO settings in Sec. \ref{MIMO_sec}, elaborating on the latest advancements in data-dependent and data-independent MIMO precoding techniques specifically designed for random ISAC signals \cite{10596930}. 

Finally, we conclude with a discussion of several open challenges and promising future research directions in sensing with random communication signals. We hope that this work will serve as a valuable contribution to the ongoing efforts to implement ISAC functionalities in the forthcoming 6G networks.


\subsection*{Notations}
Throughout this paper, $\mathsf{a}$, $\mathbf{a}$, and $\mathbf{A}$ represent random scalars, random vectors, and random matrices, respectively. Their corresponding deterministic quantities are denoted by $a$, $\bm{a}$, and $\bm{A}$, respectively. The size-$N$ identity matrix is denoted by $\bm{I}_{N}$. The size-$N$ discrete Fourier transform (DFT) matrix is denoted as $\bm{F}_{N}$, with its $(m,n)$-th entry being defined as $\frac{1}{\sqrt{N}}e^{-\frac{j2\pi(m-1)(n-1)}{N}}$ with $j$ denoting the imaginary unit. The Kronecker product and Hadarmard product between matrices $\bm{A}$ and $\bm{B}$ are denoted by $\bm{A}\otimes\bm{B}$ and $\bm{A}\odot\bm{B}$, respectively. $\|\bm{x}\|_p$ denotes the $\ell_p$ norm, which represents the $\ell_2$ norm by default when the subscript is omitted. The notations $\mathbb{E}(\cdot)$ and $\operatorname{var}(\cdot)$ denote the expectation and variance of the input argument, respectively. $(\cdot)^\ast$, $(\cdot)^T$, and $(\cdot)^H$ represent the complex conjugate, transpose, and Hermitian transpose of their arguments, respectively. $\bm{a}_n$ denotes the $n$-th column of $\bm{A}$. The notation ${\operatorname{Diag}}(\cdot)$ denotes the matrix obtained by placing its arguments on the main diagonal of a square matrix. $\operatorname{Tr}(\cdot)$ stands for the trace of a square matrix, and $\operatorname{rank}(\cdot)$ stands for the rank of a matrix. The subscripts in the aforementioned notations may be omitted when they are clear from the context.

\section{Deterministic-Random Tradeoff in ISAC Systems: An Information-Theoretic Perspective}\label{drt_sec}
In this section, we introduce the fundamental DRT between S\&C, highlighting the need for developing dedicated signal processing techniques towards random ISAC signals. We first review the capacity-distortion (C-D) theory for the state-dependent memoryless ISAC channel, framing the S\&C tradeoff as a functional optimization problem from an information-theoretic perspective. Building on this foundation, we generalize the DRT to vector Gaussian ISAC channels by depicting the CRB-rate region.

\subsection{System Model}
\begin{figure}[!t]
	\centering
	\includegraphics[width = \columnwidth]{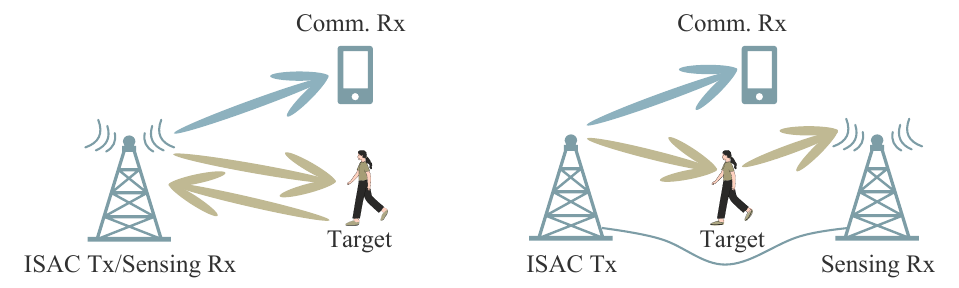}
	\caption{The P2P ISAC model: An ISAC Tx transmits a unified signal to sense targets while communicating with a communication Rx. A dedicated sensing Rx is either collocated with the ISAC Tx (monostatic mode), or placed separately but connected with the ISAC Tx through a wired link (cooperative bistatic mode).}
\label{fig:CRB_rate scenarios}
\end{figure}

We consider a P2P ISAC system, as shown in Fig. \ref{fig:CRB_rate scenarios}, where the ISAC transmitter (Tx) emits a unified signal to both sense targets and transmit information to a communication receiver (Rx). Simultaneously, a dedicated sensing receiver (Rx) is either collocated with the ISAC Tx (monostatic mode) or placed separately and connected to the Tx via a wired link (cooperative bistatic mode). In both scenarios, the sensing Rx has full knowledge of the transmitted ISAC signal, while the communication Rx does not, as the ISAC signal carries information intended for the communication user. Specifically, the primary objectives of S\&C subsystems are:
\begin{itemize}
    \item \textbf{Sensing}: Detecting the presence of targets and accurately estimating key parameters such as delay, Doppler, and angle by processing echo signals reflected from targets.
    \item \textbf{Communication:} Decoding information bits transmitted by the ISAC Tx via processing the received signal output from the communication channel.
\end{itemize}


In this model, communication performance is typically measured by the achievable rate or channel capacity, which is omitted here for brevity. The sensing performance can be characterized by the mean squared error (MSE) for estimation tasks, and detection and false-alarm probabilities for detection tasks. These metrics are explained in further detail below.
\begin{itemize}
    \item \textbf{Estimation Metrics}: Estimation accuracy can be evaluated by the difference between the ground truth and estimated value. Let $\mathsf{h}$ be the ground truth of sensing parameters, such as delay, Doppler, or angle, with $\hat{\mathsf{h}}$ denoting their estimates. The estimation error is quantified by the MSE $\epsilon$, defined as
    \begin{equation}\label{MSE}
        \epsilon: =  \mathbb{E}\left(\left|\mathsf{h} - \hat{\mathsf{h}}\right|^2\right),
    \end{equation}
    where the expectation is taken over both $\mathsf{h}$ and $\hat{\mathsf{h}}$, given their potential randomness.
    \item \textbf{Detection Metrics}: The detection problem, in its simplest form, is commonly framed as a binary hypothesis testing problem, where $\mathcal{H}_0$ hypothesis stands for the case that the sensing Rx detects only the noise, and $\mathcal{H}_1$ hypothesis signifies the situation that the sensing Rx receives target return plus noise. Accordingly, the detection probability $P_{\rm D}$ is the probability that, when the target is present, the sensing Rx correctly chooses $\mathcal{H}_1$. The false alarm probability, on the other hand, is the probability that the sensing Rx erroneously selects $\mathcal{H}_1$ when the target is absent. These probabilities can be expressed as:
    \begin{equation}
        P_{\rm D} = \Pr\left(\mathcal{H}_1|\mathcal{H}_1\right),\quad P_{\rm FA} = \Pr\left(\mathcal{H}_1|\mathcal{H}_0\right).
    \end{equation}
\end{itemize}

Notably, both detection and estimation metrics may be unified as a generic \textit{distortion measure} in the context of information theory, which is defined by a bounded distance function $d(\mathsf{h}, \hat{\mathsf{h}})$. For estimation tasks, a common distortion function is the Euclidean distance, $d(\mathsf{h}, \hat{\mathsf{h}}) = |\mathsf{h} - \hat{\mathsf{h}}|^2$, which induces the MSE in \eqref{MSE}. For detection tasks, one may define $\mathsf{h} \in \{0, 1\}$ as a binary variable indicating the presence or absence of a target, and use the Hamming distance $d(\mathsf{h}, \hat{\mathsf{h}}) = \mathsf{h} \oplus \hat{\mathsf{h}}$ as the distortion metric. Accordingly, the expected distortion in this case can be written as \cite{10206462}:
\begin{equation}\label{eq_distortion}
\begin{gathered}
  \mathbb{E}\left\{ \mathsf{h}\oplus\hat{\mathsf{h}} \right\} =  \hfill \\  \left( {1 \oplus 1} \right)\Pr \left( {\hat{\mathsf{h}}  = 1\left| {\mathsf{h} = 1} \right.} \right)  + \left( {0 \oplus 0} \right)\Pr \left( {\hat{\mathsf{h}}  = 0\left| {\mathsf{h}  = 0} \right.} \right) \hfill \\
   + \left( {1 \oplus 0} \right)\Pr \left( {\hat{\mathsf{h}}  = 1\left| {\mathsf{h}  = 0} \right.} \right)  + \left( {0 \oplus 1} \right)\Pr \left( {\hat{\mathsf{h}}  = 0\left| {\mathsf{h}  = 1} \right.} \right) \hfill \\
= 1 - {P_{\rm D}} + {P_{\rm FA}}, \hfill \\ 
\end{gathered}
\end{equation}
Under the Neyman-Pearson criterion \cite{kay1998fundamentals2}, where $P_{\rm FA}$ is fixed, minimizing the average Hamming distortion in \eqref{eq_distortion} leads to the maximization of the detection probability. 

Next, we introduce the C-D theory, using the generic distortion measure as a KPI for sensing without indicating the specific estimation or detection tasks. It is important to note that the C-D theory serves as an information-theoretic abstraction of the concrete ISAC transmission model described in Fig. \ref{fig:CRB_rate scenarios}, with an aim to characterize the achievable performance boundary for both S\&C functions under any ISAC channel models and sensing tasks, including both estimation and detection.

\begin{figure}[!t]
	\centering
	\includegraphics[width = \columnwidth]{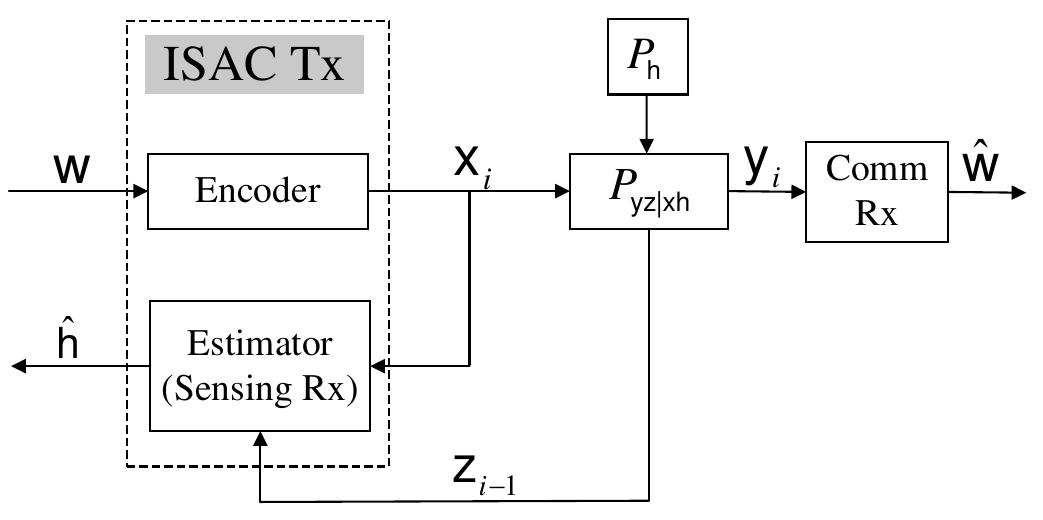}
	\caption{An information-theoretic model for the P2P monostatic ISAC system.}
    \label{ISAC_IT_Model}
\end{figure} 

\subsection{Capacity-Distortion Theory}\label{drt_sec_CD}
\subsubsection{Information-Theoretic Model for Monostatic ISAC}
The shared use of wireless resources in ISAC systems creates an inherent tradeoff between the performance of S\&C functions. The C-D theory was dedicated to analyzing such a tradeoff and the corresponding fundamental limits in a unified framework \cite{liuan2022survey,9785593}. This theory extends Shannon's rate-distortion framework, which was originally conceived for lossy data compression \cite{cover1999elements}, to the ISAC system. To elaborate, consider the P2P monostatic ISAC system as an example, where the ISAC channel is modeled as a memoryless state-dependent delayed-feedback channel \cite{8437621,8849242,9457571,9785593}, subject to a channel law, i.e., a conditional distribution $P_{\mathsf{y}\mathsf{z}|\mathsf{x}\mathsf{h}}$. Here, $\mathsf{x}$ represents the input ISAC signal, $\mathsf{h}$ is the target parameter of interest, which could be delay, Doppler, angle, or even a binary variable indicating the presence/absence of targets, and is also referred to as a ``sensing channel state''. Moreover, $\mathsf{y}$ is the signal received at the communication Rx, and $\mathsf{z}$ stands for the target return, modeled as a delayed feedback of the ISAC channel.

As shown in the information-theoretic model in Fig. \ref{ISAC_IT_Model}, the ISAC Tx consists of an encoder and an estimator (collocated sensing Rx). The encoder wishes to communicate a message $\mathsf{w}$ to the communication receiver (Rx), while simultaneously detecting/estimating the target parameter $\mathsf{h}$ by relying on the returned echo of the channel. At any time slot $i$, the channel output $\mathsf{y}_i$ and feedback $\mathsf{z}_i$ are generated based on the channel law $P_{\mathsf{y}\mathsf{z}|\mathsf{x}\mathsf{h}}(y_i, z_i |x_i, h_i)$, given the input $\mathsf{x}_i = x_i$ and state realization $\mathsf{h}_i = h_i$, which are drawn from the finite sets $\mathcal{Y}, \mathcal{Z}, \mathcal{X}$, and $\mathcal{H}$. The parameter $\mathsf{h}_i$, for all $i$, is assumed to be i.i.d. following the distribution $P_{\mathsf{h}}$.

Upon relying on the above model, a $\left(2^{nR}, n\right)$ ISAC coding scheme consists of \cite{9785593}:
\begin{itemize}
    \item A message set $\mathcal{W}$ of size at least $2^{nR}$;
    \item A series of encoding functions $\varPhi_i: \mathcal{W} \times \mathcal{Z}^{i-1} \to \mathcal{X}$, for each $i = 1, 2, \ldots, n$, which generate the channel input (ISAC signal) based on the message and past feedback, given by $\mathsf{x}_i = \varPhi_i\left(\mathsf{w},\mathsf{z}^{i-1}\right)$;
    \item A decoding function $\psi: \mathcal{H}^n \times \mathcal{Y}^n \to \mathcal{W}$, designed to recover the transmitted message using both the sequence of communication outputs $\mathsf{y}$ and the sensing channel state $\mathsf{h}$. The decoded message may be expressed as ${\hat{\mathsf{w}}} = \psi\left(\mathsf{h}^n, \mathsf{y}^n\right)$;
    \item A state estimator $\hat{h}: \mathcal{X}^n \times \mathcal{Z}^n \to {\hat{\mathcal{H}}}^n$, where ${\hat{\mathcal{H}}}^n$ denotes the finite set of reconstructed state values. This estimator uses both the transmitted ISAC signal and the feedback to recover the actual state sequence, namely, ${\hat{\mathsf{h}}}^n = \hat{h}\left(\mathsf{x}^n, \mathsf{z}^n\right)$.
\end{itemize}

\subsubsection{S\&C Performance Evaluation}
To evaluate the sensing performance of the presented ISAC system, we use the average per-block distortion by relying on a certain distortion function $d(\mathsf{h}, \hat{\mathsf{h}})$, e.g., Euclidean distance for estimation or Hamming distance for detection. This is defined as:
\begin{equation}
    \Delta^{(n)}: = \mathbb{E}\left\{d(\mathsf{h}^n, {\hat{\mathsf{h}}}^n)\right\} = \frac{1}{n}\sum\limits_{i=1}^n\mathbb{E}\left\{d(\mathsf{h}_i, {\hat{\mathsf{h}}}_i)\right\}.
\end{equation}
Thanks to the memoryless nature of the ISAC channel, it is provable that the optimal estimator ${\hat h}^\star$ is single-letterized, which minimizes the expected posterior distortion, given by
\begin{equation}
    {\hat h}^\star(x,z): = \arg\mathop{\min}\limits_{h^\prime\in\hat{\mathcal{H}}}\sum_{h\in{\mathcal{H}}}P_{\mathsf{h}|\mathsf{x}\mathsf{z}}(h|x,z)d(h,h^\prime).
\end{equation}
This yields an estimation error:
\begin{equation}\label{general_cost}
    e(x) = \mathbb{E}\left\{d[\mathsf{h},{\hat h}^\star(\mathsf{x},\mathsf{z})]|\mathsf{x} = x\right\},
\end{equation}
which acts as the sensing cost function for a given instance of ISAC signal $\mathsf{x}$. For communication performance, on the other hand, we consider the average error probability of the transmitted message:
\begin{equation}
  P_e^{(n)}: = \frac{1}{2^{nR}}\sum\limits_{i=1}^{2^{nR}}\Pr(\hat{\mathsf{w}}\ne i|\mathsf{w}=i).  
\end{equation}
To model the constraint on wireless resources (e.g., transmission power or bandwidth), we introduce a cost function $b(\cdot)$ for the ISAC signal $\mathsf{x}$, defined by
\begin{equation}
    \mathbb{E}\left\{b(\mathsf{x}^n)\right\} = \frac{1}{n}\sum\limits_{i=1}^n \mathbb{E}\left\{b(\mathsf{x}_i)\right\}.
\end{equation}
Given a resource budget $B$, a rate-distortion-cost tuple $\left\{R,D,B\right\}$ is deemed achievable if there exists a $\left(2^{nR}, n\right)$ ISAC coding scheme, such that \cite{9785593}
\begin{subequations}
\begin{align}
& \mathop {\lim }\limits_{n \to \infty } P_e^{(n)} = 0, \\
& \mathop {\lim }\limits_{n \to \infty } \Delta^{(n)} \le D,\\
& \mathop {\lim }\limits_{n \to \infty } \mathbb{E}\left\{b(\mathsf{x}^n)\right\} \le B.
\end{align}
\end{subequations}
Under the above framework, the C-D tradeoff of an ISAC system is defined as
\begin{equation}\label{CD_operational_meaning}
    C_B(D) = \operatorname{sup}\left\{R|\left\{R,D,B\right\} \text{is achievable}\right\}.
\end{equation}

Although the above definition clarifies the operational meaning of the C-D tradeoff, it is often challenging to directly compute $C_B(D)$ in a tractable manner. To that end, the following information-theoretic C-D function, which maximizes the conditional mutual information (MI) over the ISAC signal distribution $P_{\mathsf{x}}(x)$, was proved to be equivalent to \eqref{CD_operational_meaning} \cite{9785593}:
\begin{subequations}\label{CD_IT_meaning}
\begin{align}
    C_B(D) = & \arg\mathop{\max}\limits_{P_{\mathsf{x}}(x)} I\left(\mathsf{x};\mathsf{y}|\mathsf{h}\right) \label{comm_MI}\\
    &\;{\rm s.t.}\;\; \sum_{x\in\mathcal{X}}P_{\mathsf{x}}(x)e(x) \le D,\label{sensing_distortion}\\ &\quad\quad\sum_{x\in\mathcal{X}}P_{\mathsf{x}}(x)b(x) \le B,\label{resource_cost}
\end{align}    
\end{subequations}
where $I\left(\mathsf{x};\mathsf{y}|\mathsf{h}\right)$ stands for the conditional MI between the input $\mathsf{x}$ and output $\mathsf{y}$ conditioned on the channel $\mathsf{h}$, and \eqref{sensing_distortion} and \eqref{resource_cost} represent the constraints on the average sensing distortion and resource cost, respectively. For a given channel law $P_{\mathsf{y}\mathsf{z}|\mathsf{x}\mathsf{h}}$, the functional optimization problem \eqref{CD_IT_meaning} may be solved in an iterative manner via the celebrated Blahut-Arimoto (BA) algorithm \cite{1054753,1054855}. 

\subsubsection{Example}
We consider a scalar linear Gaussian channel $P_{\mathsf{y}\mathsf{z}|\mathsf{x}\mathsf{h}}$, with the following input-output relationship \cite{8437621}:
\begin{align}\label{perfect_feedback}
    \mathsf{y} = \mathsf{h}\mathsf{x} + \mathsf{n}, \quad \mathsf{z} = \mathsf{y},
\end{align}
where the channel coefficient $\mathsf{h}\in\mathbb{R}$, obeying the Gaussian distribution, is the parameter to be estimated by the ISAC signal $\mathsf{x}\in\mathbb{R}$, and $\mathsf{n}\in\mathbb{R}$ is the additive white Gaussian noise (AWGN) with zero mean and unit variance. The resource budget is the transmit power, which is set as $\mathbb{E}(|\mathsf{x}|^2)\le B = 10$. Here, the communication Rx wishes to recover $\mathsf{x}$ upon receiving $\mathsf{y}$, whereas the sensing Rx wishes to estimate $\mathsf{h}$ by observing the echo signal $\mathsf{z}$ with a minimum MSE (MMSE) estimator. We note here that the model \eqref{perfect_feedback} assumes perfect feedback. This is a toy model simply to illustrate the C-D tradeoff, which may not fully align with the realistic ISAC transmission scenario elaborated in later sections.

\begin{figure}
\centering
\subfloat[The capacity-distortion boundary.]{
\centering
\includegraphics[width=.3\textwidth]{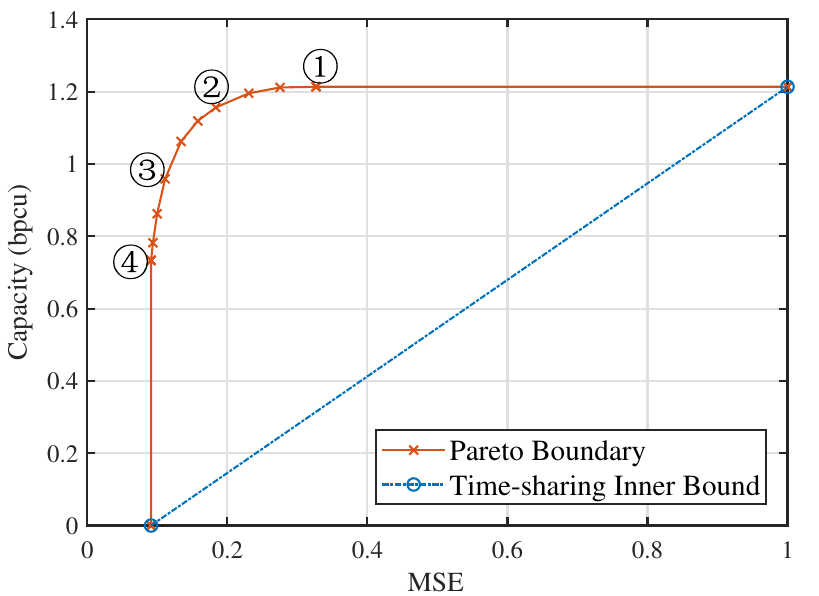}
}\\
\subfloat[$P_{\mathsf{x}}(x)$ at \textcircled{1}.]{
\includegraphics[width=.2\textwidth]{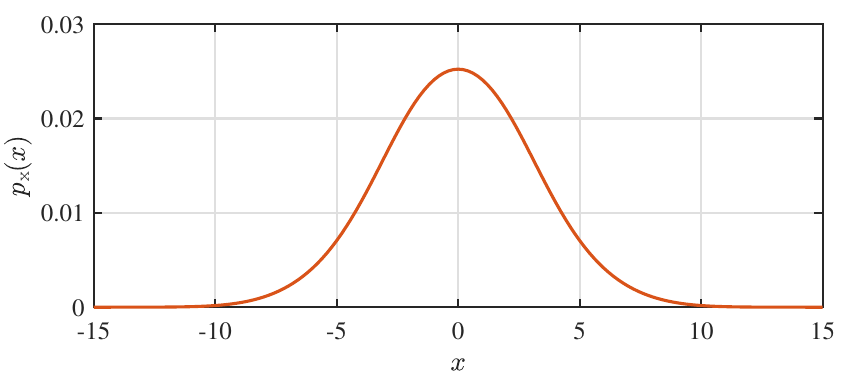}
}
\subfloat[$P_{\mathsf{x}}(x)$ at \textcircled{2}.]{
\includegraphics[width=.2\textwidth]{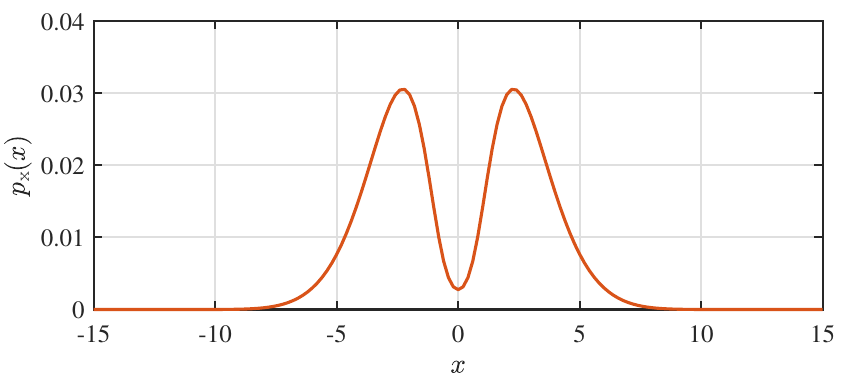}
}\\
\subfloat[$P_{\mathsf{x}}(x)$ at \textcircled{3}.]{
\includegraphics[width=.2\textwidth]{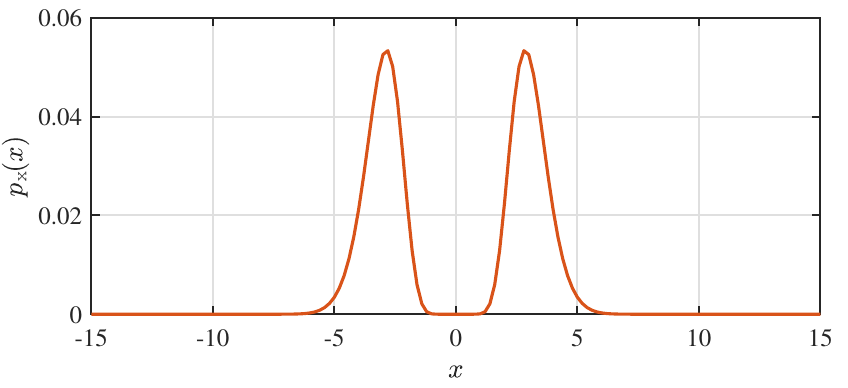}
}
\subfloat[$P_{\mathsf{x}}(x)$ at \textcircled{4}.]{
\includegraphics[width=.2\textwidth]{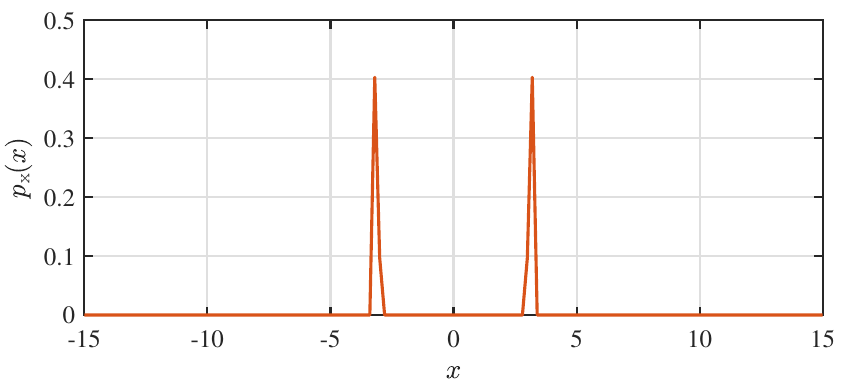}
}
\caption{The C-D tradeoff boundary of the real-valued scalar Gaussian channel scenario with $B=10$, as well as the Pareto-optimal input distributions $P_{\mathsf{x}}(x)$ along the boundary.}
\label{fig:cd_siso_distribution}
\end{figure}

Fig. \ref{fig:cd_siso_distribution} portrays the C-D tradeoff result as well as corresponding optimal ISAC signal distributions under the above model, through numerically solving problem \eqref{CD_IT_meaning}. It is observed that at points \textcircled{1} and \textcircled{4}, the optimal input distributions of $\mathsf{x}$ are the Gaussian and BPSK constellations, respectively, corresponding to the communication-optimal and sensing-optimal performance. Along the C-D tradeoff curve, $P_{\mathsf{x}}(x)$ smoothly evolves from Gaussian to BPSK, indicating a gradual reduction in the randomness of the ISAC signal \cite{8437621}.

\subsection{CRB-Rate Region for Vector Gaussian ISAC Channels}\label{sec_2b}
\subsubsection{Generic Framework}
The C-D theory presented above provides clear evidence of the DRT between S\&C in terms of the input distribution of the ISAC channel. A natural question arises as to whether this tradeoff holds in more complex scenarios, such as MIMO channels with multiple sensing parameters. To explore this, the study in \cite{Xiong_TIT} examined the CRB-rate tradeoff between S\&C in the context of the following P2P vector Gaussian ISAC channel model:
\begin{subequations}\label{vector_Gaussian_model}
\begin{align}
    &\mathbf{Y}_s = \mathbf{H}_s({\bm \upeta})\mathbf{X} + \mathbf{Z}_s \label{sensing_model}\\
    &\mathbf{Y}_c = \mathbf{H}_c\mathbf{X} + \mathbf{Z}_c,\label{comm_model}
\end{align}   
\end{subequations}
where $\mathbf{X}\in\mathbb{C}^{N_t\times N}$ denotes the dual-functional ISAC waveform emitted from the ISAC Tx, $\mathbf{Y}_s\in\mathbb{C}^{N_s\times N}$ and $\mathbf{Y}_c\in\mathbb{C}^{N_c\times N}$ are signal matrices received at the S\&C Rxs, $\mathbf{H}_s\in\mathbb{C}^{N_s\times N_t}$ and $\mathbf{H}_c\in\mathbb{C}^{N_c\times N_t}$ represent the S\&C channel matrices, while $\mathbf{Z}_s \in \mathbb{C}^{N_s\times N}$ and $\mathbf{Z}_c \in \mathbb{C}^{N_c\times N}$ are white Gaussian noise matrices, with each entry being i.i.d. and following $\mathcal{CN}\left(0,\sigma_s^2\right)$ and $\mathcal{CN}\left(0,\sigma_c^2\right)$, respectively. Moreover, the sensing channel $\mathbf{H}_s$ is assumed to be a function of the sensing parameter vector ${\bm \upeta}\in\mathbb{R}^K$, which has a prior distribution $P_{\bm{\upeta}}$ and may include delay, Doppler, and angle parameters of one or more targets. The two channels may be correlated with each other to a certain degree, depending on the specific ISAC scenarios.

The model in \eqref{vector_Gaussian_model} is generic and applicable to both monostatic and cooperative bi-static ISAC systems, as illustrated in Fig. \ref{fig:CRB_rate scenarios}. In this configuration, an ISAC Tx equipped with $N_t$ antennas emits a dual-functional ISAC signal $\mathbf{X}$ over $N$ consecutive time slots, enabling communication with an $N_c$-antenna Rx while simultaneously sensing the targets. The echo signal $\mathbf{Y}_s$ is then received at an $N_s$-antenna sensing Rx. We therefore model $\mathbf{X}$ as a random matrix governed by a distribution $P_{\mathbf{X}}\left({\bm X}\right)$, with its realization fully known to the sensing Rx but unknown to the communication Rx.

Since \eqref{comm_model} represents a P2P MIMO communication channel, the achievable communication rate is fully characterized by the input-output MI conditioned on the channel $\mathbf{H}_c$, namely, $I(\mathbf{X};\mathbf{Y}|\mathbf{H}_c)$, which is a generalization of its scalar counterpart \eqref{comm_MI}. However, unlike conventional radar systems, the sensing CRB is not straightforwardly characterized due to the randomness in the probing waveform $\mathbf{X}$. To address this issue, one may treat $\mathbf{X}$ as a \textit{random but known} nuisance parameter, and resort to the Miller-Chang bound (MCB) as a potential solution \cite{1055879}. In particular, the MCB is obtained by computing the CRB for a given instance of $\mathbf{X}$, and then taking the expectation over $\mathbf{X}$. For any weakly unbiased estimate $\hat{{\bm\upeta}}$, the resulting MSE is lower-bounded by the Bayesian MCB as follows:
\begin{equation}\label{MCB}
    \varepsilon: = \mathbb{E}\left\{\left\|\hat{{\bm\upeta}} - {\bm\upeta}\right\|^2\right\} \ge \mathbb{E}_{\mathbf{X}}\left\{\operatorname{Tr}\left(\mathbf{M}_{{\bm\upeta}|\mathbf{X}}^{-1}\right)\right\},
\end{equation}
where $\mathbf{M}_{{\bm\upeta}|\mathbf{X}}\in\mathbb{C}^{K\times K}$ is the Bayesian Fisher Information matrix (BFIM) of $\bm\upeta$ \cite{van2004detection}, which can be equivalently expressed as an affine map of the sample covariance matrix $\mathbf{R}_{\mathbf{X}}: = N^{-1}\mathbf{X}\mathbf{X}^H$ in the following form \cite{Xiong_TIT}:
\begin{align}
    &\nonumber\mathbf{M}_{{\bm\upeta}|\mathbf{X}} = {\bm\varPhi}(\mathbf{R}_{\mathbf{X}}):= \\
    &\frac{N}{\sigma_s^2}\left(\sum_{n = 1}^{r_1}\bm{H}_{1,n}\mathbf{R}_{\mathbf{X}}^T\bm{H}_{1,n}^H + \sum_{m = 1}^{r_2}\bm{H}_{2,m}\mathbf{R}_{\mathbf{X}}\bm{H}_{2,m}^H + \bm{M}_P\right),
\end{align}
with the term $\bm{M}_P$ representing the prior FIM contributed by $P_{\bm\upeta}$. Additionally, the matrices $\bm{H}_{1,n}$ and $\bm{H}_{2,m}$ are partitioned from the Jacobian matrix $\frac{\partial \operatorname{vec}(\mathbf{H}_{s}^{\ast})}{\partial \bm{\upeta}}$.

\begin{figure}[!t]
	\centering
	\includegraphics[width = \columnwidth]{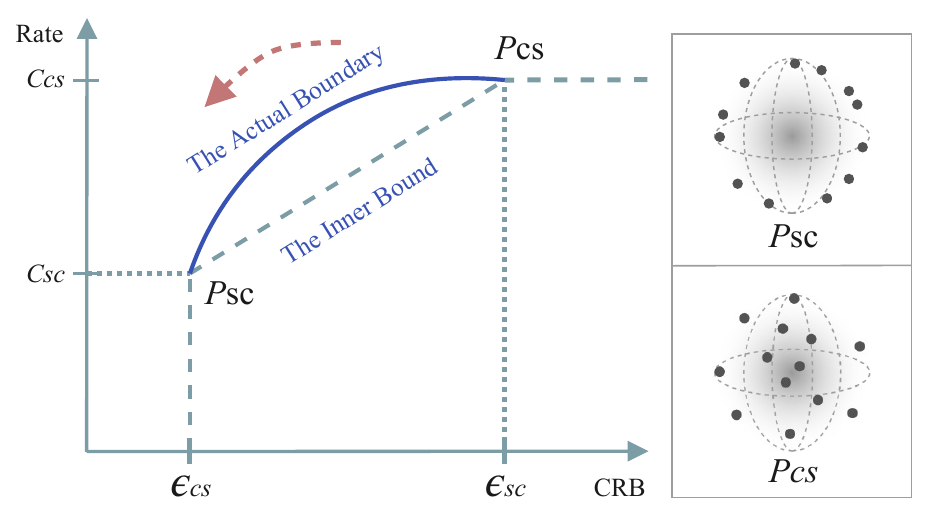}
	\caption{CRB-Rate tradeoff for a P2P monostatic ISAC system.}
    \label{CRB_rate_tradeoff}
\end{figure}

The Bayesian MCB in \eqref{MCB} may be achieved by the \textit{maximum a posterior (MAP)} estimator at the high SNR regime \cite{van2004detection}. By comparing \eqref{MCB} and \eqref{sensing_distortion}, it becomes evident that the MCB serves as an equivalent average sensing cost function of the random ISAC signal $\mathbf{X}$, even though it is not a traditional ``distortion'' metric. Consequently, the CRB-rate tradeoff in the ISAC system can be framed as the following Pareto optimization problem:
\begin{subequations}\label{opt_problem}
\begin{align}
&\min_{P_{\mathbf{X}}(\bm{X})} \alpha\mathbb{E}\left\{\operatorname{Tr}{\left[\left(\bm{\varPhi}({\mathbf{R}}_{\mathbf{X}})\right)^{-1}\right]}\right\} - (1-\alpha)I(\mathbf{X};\mathbf{Y}_{c}|\mathbf{H}_{c}) \label{obj_function_CRB}\\
&\;\;\;{\rm s.t.}\;\;\mathbb{E}\left\{\operatorname{Tr}\left({{\mathbf{R}}_{\mathbf{X}}}\right)\right\}=P_{T}, \label{opt_constraints}
\end{align}
\end{subequations}
where $P_T$ stands for the average transmit power budget, and $\alpha\in\left[0,1\right]$ is a parameter that balances the tradeoff between S\&C objectives. As illustrated in Fig. \ref{CRB_rate_tradeoff}, adjusting $\alpha$ from $0$ to $1$ traces out the Pareto boundary of the two objectives, capturing the achievable CRB-rate pairs and thus defining the CRB-rate region of the ISAC system. 

Despite the convexity of \eqref{opt_problem} with respect to $P_{\mathbf{X}}$, obtaining the complete CRB-rate boundary is computationally challenging due to the infinite dimensionality of $P_{\mathbf{X}}$. Therefore, our focus shifts to the two extreme points $P_{\rm CS}$ and $P_{\rm SC}$, representing the optimal ISAC signal distributions for communication and sensing, respectively. Notably, the line segment connecting these two points forms a time-sharing inner bound for the CRB-rate region.

\subsubsection{S\&C Performance at $P_{\rm CS}$}
Given the linear Gaussian model in \eqref{comm_model}, the communication rate at $P_{\rm CS}$ is maximized by a Gaussian input distribution, where each column of $\mathbf{X}$  is independently drawn from $\mathcal{CN}(\mathbf{0},\widetilde{\bm{R}}_{\mathbf{X}}^{\rm CS})$, with a statistical covariance matrix defined by
\begin{equation}
    \widetilde{\bm{R}}_{\mathbf{X}}^{\rm CS} = \bm{U}_c\bm{\varLambda}_{\rm CS}\bm{U}_c^H,
\end{equation}
where $\bm{U}_c$ contains the right singular vectors of $\mathbf{H}_c$, and $\bm{\varLambda}_{\rm CS}$ is a diagonal matrix with eigenvalues determined by the water-filling power allocation approach. Consequently, the optimal ISAC signal structure at $P_{\rm CS}$ is given by
\begin{equation}\label{opt_X_CS}
    \mathbf{X}^{\rm CS} = \bm{U}_c\bm{\varLambda}_{\rm CS}^{1/2}\mathbf{D},
\end{equation}
where $\mathbf{D}$ is an information-bearing random matrix containing i.i.d. entries following $\mathcal{CN}(0,1)$. The resulting achievable rate at $P_{\rm CS}$ can then be expressed by the renowned Shannon capacity formula as
\begin{equation}
    C_{\rm CS} = \mathbb{E}\left\{\log\det\left(\bm{I} + \sigma_c^{-2}\mathbf{H}_c\widetilde{\bm{R}}_{\mathbf{X}}^{\rm CS}\mathbf{H}_c^H\right)\right\}.
\end{equation}

In contrast to the communication rate that depends on the statistical covariance matrix of $\mathbf{X}$, the sensing CRB is determined by the sample covariance matrix $\mathbf{R}_{\mathbf{X}}$. The latter follows a complex Wishart distribution due to the Gaussian-distributed $\mathbf{X}^{\rm CS}$ at $P_{\rm CS}$. Unfortunately, the CRB at $P_{\rm CS}$, which involves calculating the expectation of a highly nonlinear function of $\mathbf{R}_{\mathbf{X}}$, is unlikely to be expressed in a closed form. To that end, one may resort to establishing the lower- and upper-bounds of the sensing CRB $\epsilon_{\rm CS}$ as follows \cite{Xiong_TIT}:
\begin{equation}\label{Gaussian_CRB}
    \operatorname{Tr}{\left\{\left[\bm{\varPhi}(\widetilde{\bm{R}}_{\mathbf{X}}^{\rm CS})\right]^{-1}\right\}}\le \epsilon_{\rm CS} \le \frac{N\cdot\operatorname{Tr}{\left\{\left[\bm{\varPhi}(\widetilde{\bm{R}}_{\mathbf{X}}^{\rm CS})\right]^{-1}\right\}}}{N-\min\left\{K,\operatorname{rank}(\widetilde{\bm{R}}_{\mathbf{X}}^{\rm CS})\right\}}.
\end{equation}
Observe that the two bounds become identical when $N\to\infty$.

\subsubsection{S\&C Performance at $P_{\rm SC}$}
Evaluating S\&C performance at point $P_{\rm SC}$ is generally more challenging than at $P_{\rm CS}$, as neither the optimal ISAC signal distribution nor the achievable communication rate is explicitly determined at this stage. Denoting the CRB at $P_{\rm SC}$ as $\epsilon_{\rm SC}$, we can characterize it using the Jensen's inequality:
\begin{align}\label{Jensen_Psc}
    &\nonumber\mathbb{E}\left\{\operatorname{Tr}{\left[\left(\bm{\varPhi}({\mathbf{R}}_{\mathbf{X}})\right)^{-1}\right]}\right\} \ge \operatorname{Tr}\left\{\left(\bm{\varPhi}\left[\mathbb{E}\left({\mathbf{R}}_{\mathbf{X}}\right)\right]\right)^{-1}\right\} \\
    &\ge \operatorname{Tr}\left\{\left[\bm{\varPhi}\left(\widetilde{\bm{R}}_{\mathbf{X}}^{\rm SC}\right)\right]^{-1}\right\} := \epsilon_{\rm SC},
\end{align}
where $\widetilde{\bm{R}}_{\mathbf{X}}^{\rm SC}$ is the optimal covariance matrix at $P_{\rm SC}$, attained by solving the deterministic CRB minimization problem:
\begin{equation}\label{BCRB_opt}
{{\widetilde{\bm{R}}_{\mathbf{X}}^{\rm{SC}}}} = \mathop {\arg \min }\limits_{{\widetilde{\bm{R}}} \succeq {\mathbf{0}},\;{\widetilde{\bm{R}}} = {{\widetilde{\bm{R}}}^{H}}} \;{\operatorname{Tr}}\left\{ \left[{{{\bm{\varPhi }}}( {\widetilde{\bm{R}}} )}\right]^{ - 1}\right\}\;\;\operatorname{s.t.}\;{\operatorname{Tr}}( {\widetilde{\bm{R}}}) \le {P_{T}}.
\end{equation}
The equality in \eqref{Jensen_Psc} holds if and only if
\begin{equation} \label{eqn:21}
\frac{1}{N}\mathbf{X}\mathbf{X}^H = {\mathbf{R}}_{\mathbf{X}} = \mathbb{E}\left({\mathbf{R}}_{\mathbf{X}}\right) = {{\widetilde{\bm{R}}_{\mathbf{X}}^{\rm{SC}}}}, 
\end{equation}
indicating that $\mathbf{R}_\mathbf{X}$ becomes deterministic at $P_{\rm SC}$\footnote{It is worth mentioning a special case when $N\to\infty$, in which $\mathbf{R}_\mathbf{X}$ becomes deterministic at $P_{\rm SC}$, such that the equality in \eqref{Jensen_Psc} and \eqref{eqn:21} both hold. In this case, due to the deterministic property of $\mathbf{R}_\mathbf{X}$, the Gaussian signaling becomes optimal for both S\&C functions, and we can optimize the covariance matrix $\mathbf{R}_\mathbf{X}$ to characterize the complete Pareto boundary between $P_{\rm SC}$ and $P_{\rm SC}$; see, e.g., \cite{10217169,10251151} for examples when $\mathbf{H}_s({\bm \upeta})$ and $\mathbf{H}_c$ are deterministic (instead of random in our context). This, however, only serves as an S\&C performance upper bound for the general case with $L$ being finite in general.}. This suggests the optimal ISAC signal structure at $P_{\rm SC}$ as follows:
\begin{equation}\label{opt_X_SC}
    \mathbf{X}^{\rm SC} = \sqrt{N}({{\widetilde{\bm{R}}_{\mathbf{X}}^{\rm{SC}}}})^{1/2}\mathbf{Q} = \sqrt{N}\bm{U}_s\bm{\varLambda}_{\rm SC}^{1/2}\mathbf{Q},
\end{equation}
where $\bm{U}_s$ and $\bm{\varLambda}_{\rm SC}$ are the eigenvectors and eigenvalues of ${{\widetilde{\bm{R}}_{\mathbf{X}}^{\rm{SC}}}}$, and $\mathbf{Q} \in \mathbb{C}^{\operatorname{rank}(\bm{\varLambda}_{\rm SC})\times N}$ is a semi-unitary matrix (i.e., $\mathbf{Q}\mathbf{Q}^H = \bm{I}$) that carries communication data, which belongs to the Stiefel manifold. The communication degrees of freedom (DoFs) are constrained to $\mathbf{Q}$ due to the deterministic nature of ${{\widetilde{\bm{R}}_{\mathbf{X}}^{\rm{SC}}}}$. Under this configuration, the achievable communication rate can be expressed as
\begin{equation}\label{BCRB_opt_rate}
C_{\rm SC} = \mathop {\arg \max }\limits_{P_{\mathbf{Q}}(\bm{Q})}\;\;I\left(\mathbf{Q};\mathbf{Y}_c|\mathbf{H}_c\right)\;\;\operatorname{s.t.}\;\mathbf{Q}\mathbf{Q}^H = \bm{I}.
\end{equation}
Deriving an explicit form of $C_{\rm SC}$ is challenging; however, as shown in \cite{Xiong_TIT}, the asymptotic rate at $P_{\rm SC}$ in the high SNR regime is given by:
\begin{align}\label{comm_dof_loss}
C_{\rm SC} = \nonumber &\mathbb{E}\Big\{\Big(1-\frac{\operatorname{rank}(\bm{\varLambda}_{\rm SC})}{2N}\Big)\log\det(\sigma_{ c}^{-2}\mathbf{H}_{c}\widetilde{\bm{R}}_{\mathbf{X}}^{\rm SC}\mathbf{H}_{c}^{H})+c_0\Big\}\\
& + O(\sigma_{c}^2),
\end{align}
where $c_0$ is irrelevant to the SNR and approaches zero as $N \to \infty$. The rate in \eqref{comm_dof_loss} may be achieved asymptotically by a uniformly distributed $\mathbf{Q}$ over the Stiefel manifold.

\subsubsection{DRT in Vector Gaussian ISAC Channels}
We now turn to the discussion of the DRT for the vector Gaussian ISAC model in \eqref{vector_Gaussian_model}, by examining the S\&C performance at two corner points. First, it is clear from the structures of \eqref{opt_X_CS} and \eqref{opt_X_SC} that the randomness level of ISAC signals decreases as the system shifts from the communication-optimal point to the sensing-optimal point. This occurs because the communication codewords transition from the i.i.d. Gaussian matrix $\mathbf{D}$ to the uniformly distributed semi-unitary matrix $\mathbf{Q}$. This tradeoff in the optimal input distribution of ISAC signals can be seen as a generalized form of the Gaussian-BPSK tradeoff in the scalar Gaussian channel, discussed earlier in the C-D theory.

Additionally, the DRT is evident in the achievable rate \eqref{BCRB_opt_rate} at $P_{\rm SC}$, where a reduction in communication DoFs leads to a rate loss compared to the Gaussian capacity $C_{\rm SC}$. Conversely, greater randomness in the ISAC signal can impair the sensing performance, as seen in the CRB. Specifically, \eqref{Gaussian_CRB} indicates that a Gaussian-distributed ISAC signal could inflate the CRB by a factor greater than $1$, in contrast to the minimum CRB achievable at $P_{\rm SC}$ by signals with deterministic sample covariance matrices.

\subsection{From DRT Theory to Random ISAC Signal Processing}
The information-theoretic insights presented above underscore the S\&C tradeoff linked to the input distribution of $\mathbf{X}$, highlighting a novel design DoF in ISAC systems and unveiling a range of research opportunities, particularly in \textit{ISAC transmission with random communication signals}. Unlike conventional radar systems that rely on deterministic signals or structured pseudo-random sequences, e.g., $m$-sequences or Gold sequences, future 6G ISAC systems must utilize communication data-carrying signals for sensing, which are endowed with \textit{randomness} due to their information-bearing nature. While this randomness enhances the communication rate, it may also introduce random fluctuations in the echo signals, potentially deteriorating the target detection and estimation performance. As such, it becomes essential to define new sensing performance metrics and to develop tailored processing techniques for random ISAC signals. 

Expanding on the theoretical basis of ISAC, we next overview the recent research progress on the characterization of achievable sensing performance of standard communication signals, and explore how to design basic functional blocks to either minimize sensing loss or enable a flexible S\&C performance tradeoff. This involves the configuration of modulation schemes, constellation mapping, pulse-shaping filters, and MIMO precoders. Other crucial aspects, such as source and channel coding, as well as sampling and quantization, represent valuable future research directions and are not covered here due to space constraints. In the following sections, we first delve into the ISAC transmission with random communication signals for single-antenna systems in Sec. \ref{comm_centric_model_sec} and Sec. \ref{waveform_design_sec}, and then extend our analysis to MIMO-ISAC systems in Sec. \ref{MIMO_sec}.


\section{Communication-Centric ISAC Transmission}\label{comm_centric_model_sec}
In this section, we investigate concrete signal processing methods of the basic P2P ISAC model in Fig. \ref{fig:CRB_rate scenarios}, by considering a single-antenna ISAC system comprising an ISAC Tx, a communication Rx, and a sensing Rx. The ISAC Tx emits a random data payload signal intended for the communication Rx. Simultaneously, the sensing Rx captures the echo of this signal reflected from multiple targets, and subsequently estimates the targets' delay parameters. Consistent with the system setting and assumptions in Sec. \ref{sec_2b}, the information symbol vector $\mathbf{s}$ is unknown to the communication Rx, but is fully known to the sensing Rx.

\begin{figure*}
\centering
\includegraphics[width=\textwidth]{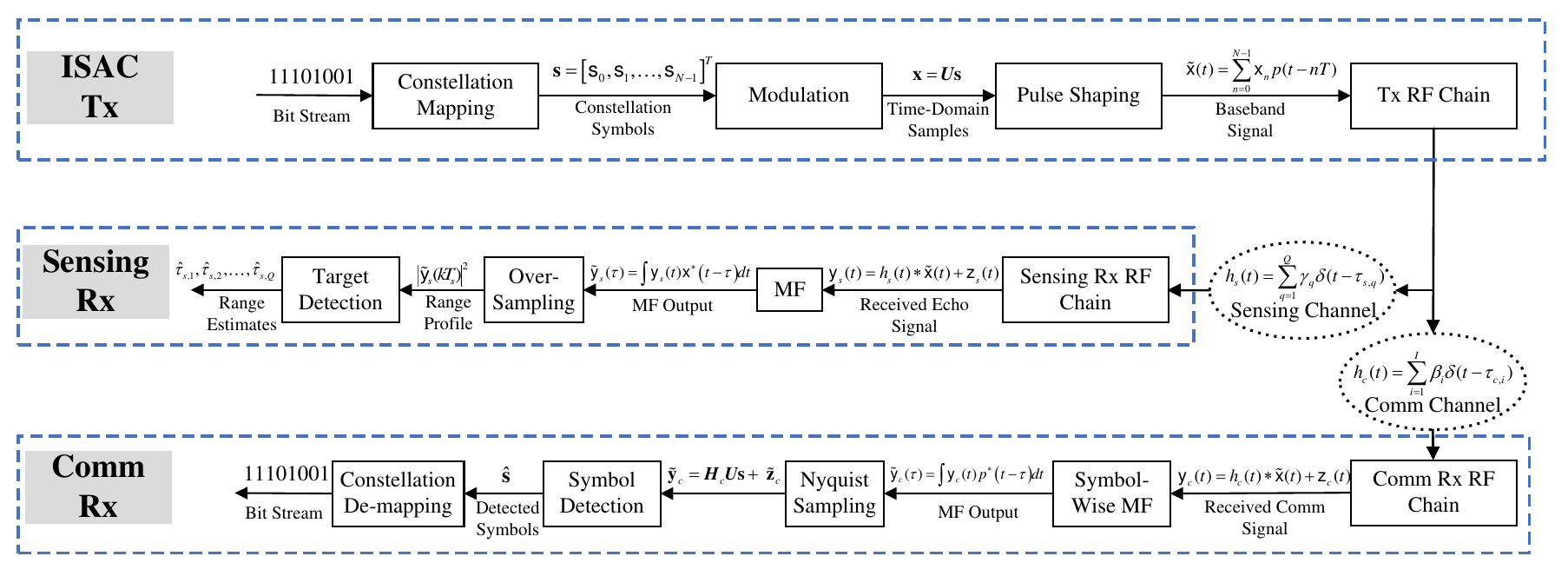}
\caption{A signal processing pipeline for P2P ISAC model in Fig. \ref{fig:CRB_rate scenarios}. The communication Rx aims to correctly detect symbols in the presence of multi-path interference, whereas the sensing Rx aims at extracting range parameters of multiple targets.}
\label{Random_SP_Pipeline}
\end{figure*}

\subsection{Communication-Centric ISAC Signal Model}
Let us consider the following generic baseband ISAC signal: 
\begin{equation}\label{baseband_ISAC_signal} 
\widetilde{\mathsf{x}}(t) = \sum\limits_{n = 0}^{N-1}\mathsf{x}_n p(t-nT), 
\end{equation} 
where $T$ stands for the symbol duration, $p(t)$ is the impulse response of a \textit{pulse shaping filter}, and $\mathbf{x} = \left[\mathsf{x}_0,\mathsf{x}_1,\ldots,\mathsf{x}_{N-1}\right]^T \in \mathbb{C}^{N\times 1}$ represents $N$ discrete time-domain samples, generated from 
\begin{equation} \label{time_domain_discrete_sample}
\mathbf{x} = \bm{U}\mathbf{s} = \sum_{n = 0}^{N-1}\mathsf{s}_n\bm{u}_n, 
\end{equation}
where $\mathbf{s} = \left[\mathsf{s}_0, \mathsf{s}_1, \ldots, \mathsf{s}_{N-1}\right]^T \in \mathbb{C}^{N\times 1}$ denotes $N$ data symbols that are independently drawn from a \textit{constellation} $\mathcal{S}$, and $\bm{U} = \left[\bm{u}_0, \bm{u}_1, \ldots, \bm{u}_{N-1}\right] \in \mathbb{U}(N)$ is an $N$-dimensional unitary matrix representing the orthonormal \textit{modulation basis} in the time domain. Note that signal model \eqref{baseband_ISAC_signal} serves as a practical realization of the information-theoretic model in Sec. \ref{drt_sec_CD}. The assumptions for these three components are elaborated as follows:

\subsubsection{Constellation} Throughout the paper, we adopt a rotationally invariant constellation $\mathcal{S}$ with zero mean, zero pseudo-variance, and unit power, defined as: 
\begin{equation} \label{regularized_constraints}
\mathbb{E}(\mathsf{s}) = 0, \quad \mathbb{E}(\mathsf{s}^2) = 0, \quad \mathbb{E}(|\mathsf{s}|^2) = 1, \quad \forall~~ \mathsf{s}\in \mathcal{S}. 
\end{equation}
We remark that most commonly employed constellations, such as PSK, QAM, and APSK, satisfy these criteria. The two exceptions are BPSK and 8-QAM, which are seldom used in modern cellular wireless networks. Additionally, the constellation does not need to be uniformly distributed; each point may be transmitted with distinct probabilities if the probabilistic constellation shaping (PCS) technique is applied, as will be discussed later.

\subsubsection{Modulation Basis} 
The modulation basis $\bm{U}\in\mathbb{U}(N)$, which carries information symbols, is often referred to as a ``waveform'' in a broader context, with $\mathbb{U}(N)$ denoting the size-$N$ unitary group. For clarity, we list several typical examples below:
\begin{itemize} 
    \item \textbf{SC Modulation:} $\bm{U} = \bm{I}_N$. In this case, the signaling basis consists solely of $N$ time-domain Kronecker-Delta functions, forming an SC signal.
    \item \textbf{OFDM Modulation:} $\bm{U} = \bm{F}_N^H$, where $\bm{F}_N\in\mathbb{C}^{N\times N}$ represents the $N$-dimensional DFT matrix. Here, the signaling basis is constructed using $N$ orthogonal sinusoidal functions, which are also Kronecker-Delta functions in the frequency domain.
    \item \textbf{CDMA Modulation:} $\bm{U} = \bm{C}_N$, where $\bm{C}_N\in\mathbb{C}^{N\times N}$ denotes the $N$-dimensional Hadamard matrix, corresponding to Walsh codes widely used in CDMA2000 \cite{995853}.
    \item \textbf{AFDM Modulation:} $\bm{U} = \bm{\varLambda}_{c_1}^H\bm{F}_N^H\bm{\varLambda}_{c_2}^H$, where $\mathbf{\varLambda}_{c} = \operatorname{Diag}(1, e^{-j2\pi c_1^2}, \ldots, e^{-j2\pi c_N^2})$. This configuration makes $\bm{U}$ an inverse discrete affine Fourier transform (IDAFT) matrix with tunable parameters $c_1$ and $c_2$, with symbols placed in the affine Fourier transform (AFT) domain \cite{9727202}.
    \item \textbf{OTFS Modulation:} $\bm{U} = \bm{F}_{N_1}^H\otimes\bm{I}_{N_2}$. In this case, the symbols are mapped to the delay-Doppler (DD) domain, where $N_1$ and $N_2$ represent the number of occupied time slots and subcarriers, respectively. Note that OTFS is inherently a 2-dimensional modulation scheme \cite{7925924}.
\end{itemize}
To facilitate efficient signal processing in the frequency domain, we assume that a cyclic prefix (CP) is added to the signal $\mathbf{x}$, which is longer than the maximum delay of the target or communication path.



\subsubsection{Pulse Shaping Filter} 
The pulse shaping filter plays a crucial role in modern wireless communication systems by eliminating inter-symbol interference (ISI) while limiting the signaling bandwidth. Here, we employ a band-limited Nyquist prototype pulse with a one-sided bandwidth $B$ and a roll-off factor $\alpha$, resulting in a symbol duration of $T = \frac{1+\alpha}{2B}$. The Nyquist pulse ensures zero ISI among symbols, which can be expressed as the following condition:
\begin{equation}
\widetilde p\left( {nT} \right) = \left\{ \begin{gathered}
  1,\;\;n = 0 \hfill \\
  0,\;\;n \ne 0 \hfill \\ 
\end{gathered}  \right.,\quad \forall~ n \in \mathbb{Z},
\end{equation}
where $\widetilde p\left(\tau\right) = \int p\left(t\right) p^\ast\left(t - \tau\right) dt$ is the ACF of $p\left(t\right)$. This translates to an equivalent frequency-domain condition, known as the \textit{folded-spectrum criterion}, given by \cite{proakis2008digital}
\begin{equation} \label{folded_spectrum}
\sum\limits_{m = -\infty}^{\infty} g\left(f + \frac{m}{T}\right) = T, 
\end{equation} 
where $g(f)$ is the Fourier transform of $\widetilde p\left(\tau\right)$, which is also the squared frequency spectrum of the pulse $p(t)$.

\subsection{Signal Processing for Communication and Sensing}
In this subsection, we elaborate on the signal processing pipeline for both S\&C, which is also illustrated in Fig. \ref{Random_SP_Pipeline}. Without loss of generality, we model both communication and sensing channels as linear time-invariant (LTI) multi-path channels containing $I$ paths and $Q$ targets, respectively. Their time-domain impulse responses can be expressed as
\begin{align}
    h_c(t) = \sum_{i = 1}^I \beta_i\delta(t - \tau_{c,i}), \quad h_s(t) = \sum_{q = 1}^Q \gamma_q\delta(t - \tau_{s,q}),
\end{align}  
where $\delta(t)$ is the Dirac-Delta function, $\beta_i$ and $\gamma_q$ denote the complex amplitudes of the $i$-th communication path and the $q$-th sensing target, respectively, with $\tau_{c,i}$ and $\tau_{s,q}$ representing the corresponding delays.
For simplicity, we omit the impact of Doppler phase shifts on the two channels here, reserving those effects for future discussions. Note that $h_c(t)$ and $h_s(t)$ may exhibit certain correlations depending on the geometric environment, indicating that communication paths and sensing targets might partially overlap. This overlap gives rise to another fundamental tradeoff in S\&C, known as the subspace tradeoff (ST). Due to space constraints, we will not delve into the details of the ST here and instead refer interested readers to \cite{10471902} for further information.

\subsubsection{Receive Signal Model}
By transmitting the ISAC signal $\widetilde{\mathsf{x}}(t)$, the received signals at the communication and sensing Rxs are expressed as
\begin{subequations}\label{signal_convolution}
\begin{align}
    \nonumber \mathsf{y}_c(t) &= h_c(t)\ast \widetilde{\mathsf{x}}(t) + \mathsf{z}_c(t)\\
    & = \sum_{i = 1}^I \beta_i\sum_{n = 0}^{N-1} \mathsf{x}_n p(t - nT - \tau_{c,i}) + \mathsf{z}_c(t), \label{comm_signal_convolution}\\
     \nonumber \mathsf{y}_s(t)& = h_s(t)\ast \widetilde{\mathsf{x}}(t) + \mathsf{z}_s(t)\\
    & = \sum_{q = 1}^Q \gamma_q\sum_{n = 0}^{N-1} \mathsf{x}_n p(t - nT - \tau_{s,q}) + \mathsf{z}_s(t), \label{sensing_signal_convolution} 
\end{align}  
\end{subequations}
where $\mathsf{z}_s(t)$ and $\mathsf{z}_s(t)$ stand for the zero-mean white Gaussian noise with variances $\sigma_c^2$ and $\sigma_s^2$, respectively. For the communication Rx, the objective is to detect the information symbols $\mathbf{s}$ embedded in $\widetilde{\mathsf{x}}(t)$ from \eqref{comm_signal_convolution}, using an estimate of the channel $h_c(t)$. In contrast, the sensing Rx aims to detect the $Q$ targets and extract corresponding delay parameters $\left\{\tau_{s,q}\right\}$ by the observation in \eqref{sensing_signal_convolution}, with prior knowledge of $\widetilde{\mathsf{x}}(t)$.

\subsubsection{Signal Processing for Communication}
Let us first discuss the signal processing procedure at the communication Rx's side. Upon receiving $\mathsf{y}_c(t)$, the communication Rx performs a symbol-wise matched-filtering (MF) by using the pulse shaping filter, leading to the following output signal \cite{proakis2008digital}:
\begin{align}
    \widetilde{\mathsf{y}}_c(\tau) =  \int {\mathsf{y}_c(t) p^\ast\left( {t - \tau } \right)} dt.
\end{align}
Sampling at $\tau = \ell T$, where $\ell\in\mathbb{Z}$, yields
\begin{equation}
    \widetilde{\mathsf{y}}_c(\ell T) = \sum_{i = 1}^I \beta_i\sum_{n = 0}^{N-1} \mathsf{x}_n \widetilde{p}_{\ell-n,\tau_{c,i}} + \widetilde{\mathsf{z}}_{c,\ell},
\end{equation}
where $\widetilde{p}_{k,\tau}: = \widetilde{p}(kT - \tau)$, and $\widetilde{\mathsf{z}}_{c,l} = \int \mathsf{z}_c(t) p^\ast(t - \ell T) dt$ represents the output noise, which remains Gaussian distributed. By defining $\widetilde{\mathbf{y}}_c$ as the discrete MF output vector, where its $(\ell+1)$-th entry corresponds to $\widetilde{\mathsf{y}}_c(\ell T)$, we obtain:
\begin{equation}\label{MF_output_comm_Rx}
    \widetilde{\mathbf{y}}_c = \sum_{i = 1}^I \beta_i\widetilde{\bm{P}}_i\mathbf{x} + \widetilde{\mathbf{z}}_{c}:= \bm{H}_c\bm{U}\mathbf{s} + \widetilde{\mathbf{z}}_{c},
\end{equation}
where $\widetilde{\mathbf{z}}_{c} = \left[\widetilde{\mathsf{z}}_{c,0}, \widetilde{\mathsf{z}}_{c,1},\ldots,\widetilde{\mathsf{z}}_{c,N-1}\right]^T$, and
\begin{equation}
   \widetilde{\bm{P}}_i = \left[ {\begin{array}{*{20}{c}}
  {{\widetilde{p}_{0,{\tau _{c,i}}}}}&{{\widetilde{p}_{ - 1,{\tau _{c,i}}}}}& \ldots &{{\widetilde{p}_{ - N + 1,{\tau _{c,i}}}}} \\ 
  {{\widetilde{p}_{1,{\tau _{c,i}}}}}&{{\widetilde{p}_{0,{\tau _{c,i}}}}}& \cdots &{{\widetilde{p}_{ - N + 2,{\tau _{c,i}}}}} \\ 
   \vdots & \vdots & \ddots & \vdots  \\ 
  {{\widetilde{p}_{N - 1,{\tau _{c,i}}}}}&{{\widetilde{p}_{N - 2,{\tau _{c,i}}}}}& \ldots &{{\widetilde{p}_{0,{\tau _{c,i}}}}} 
\end{array}} \right]
\end{equation}
is a Toeplitz matrix, making the equivalent channel matrix $\bm{H}_c$ also Toeplitz. If a CP is added to the time-domain sequence $\mathbf{x}$, then $\bm{H}_c$ becomes approximately circulant after the CP is removed from the received signal. It can be observed that ISI exists among the entries of $\mathbf{s}$ since $\bm{H}_c \bm{U}$ is not a diagonal matrix. To recover the information symbols, the ISI needs to be eliminated through channel equalization. This is typically achieved by first estimating $\bm{H}_c$ using known pilot symbols, followed by implementing an equalizer based on the estimate.

To reduce signal processing complexity, one may employ OFDM modulation by setting $\bm{U} = \bm{F}_N^H$, which yields:
\begin{align}
    \widetilde{\mathbf{y}}_c^{\text{OFDM}} &\nonumber = \bm{H}_c\bm{F}_N^H\mathbf{s} + \widetilde{\mathbf{z}}_{c} \\ &\nonumber=\sqrt{N}\bm{F}_N^H\operatorname{Diag}\left(\bm{F}_N\bm{h}_c\right)\bm{F}_N\bm{F}_N^H\mathbf{s}+ \widetilde{\mathbf{z}}_{c}\\
    &= \sqrt{N}\bm{F}_N^H\operatorname{Diag}\left(\bm{F}_N\bm{h}_c\right)\mathbf{s}+ \widetilde{\mathbf{z}}_{c} \label{OFDM_comm_processing},
\end{align}
where we utilize the property that a circulant matrix can be diagonalized by the DFT matrix \cite{Bamieh2018Circulant}, with $\bm{h}_c$ denoting the first column of $\bm{H}_c$. Accordingly, the ISI can be removed by simply performing a DFT on \eqref{OFDM_comm_processing}, resulting in:
\begin{equation}
    \bm{F}_N\widetilde{\mathbf{y}}_c^{\text{OFDM}} = \sqrt{N}\operatorname{Diag}\left(\bm{F}_N\bm{h}_c\right)\mathbf{s} + \bm{F}_N \widetilde{\mathbf{z}}_{c},
\end{equation}
which can be readily processed as $N$ parallel additive white Gaussian noise (AWGN) channels.

While OFDM minimizes the complexity of symbol detection by diagonalizing the multi-path channels, other modulation schemes such as SC, CDMA, AFDM, and OTFS can also be employed to satisfy specific application requirements, such as ensuring reliable communication in high-mobility channels.

\subsubsection{Signal Processing for Sensing}
We now turn our focus to processing the received sensing signal \eqref{sensing_signal_convolution} to extract the target delay parameters, where the first step is also to perform MF over the observed echo signal $\mathsf{y}_s(t)$. In sharp contrast to its communication counterpart, the matched filter used for sensing is the transmitted baseband signal $\widetilde{\mathsf{x}}(t)$ rather than the pulse shaping filter $p(t)$, leading to the following output \cite{levanon2004radar}:
\begin{align}\label{range_profile}
    &\widetilde{\mathsf{y}}_s(\tau) \nonumber =  \int {{\mathsf{y}}_s(t)\widetilde{\mathsf{x}}^\ast\left( {t - \tau } \right)} dt\\
    &\nonumber = \sum_{q = 1}^Q \gamma_q \int {\widetilde{\mathsf{x}}(t-\tau_{s,q})\widetilde{\mathsf{x}}^\ast\left( {t - \tau } \right)} dt + \int {\mathsf{z}_s(t)\widetilde{\mathsf{x}}^\ast\left( {t - \tau } \right)} dt\\
    & = \sum_{q = 1}^Q \gamma_q \mathsf{R}(\tau-\tau_{s,q}) + \widetilde{\mathsf{z}}_s(\tau),\quad 0\le \tau \le NT.
\end{align}
where $\mathsf{R}(\tau) = \int \widetilde{\mathsf{x}}(t) \widetilde{\mathsf{x}}^\ast(t - \tau) dt$ is the ACF of the ISAC signal $\widetilde{\mathsf{x}}(t)$, and $\widetilde{\mathsf{z}}_s(\tau)$ is the output Gaussian noise. It is worthwhile to point out that other methods, such as compressive sensing, may also be adopted to reduce sampling rates and improve resolution \cite{8498083}. In order to highlight the impact of random data over sensing performance, here we employ the most basic MF approach for ranging, while designating the investigation of more advanced algorithms as our future work.

\begin{figure}[!t]
	\centering
	\includegraphics[width = \columnwidth]{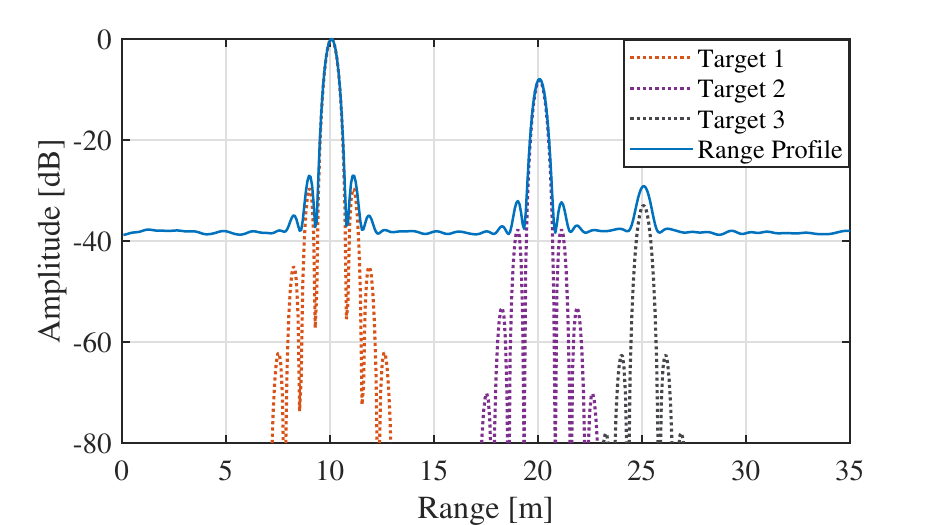}
	\caption{An example of the range profile for OFDM signal with PSK constellation, including 3 targets located at 10m, 20m, and 25m.}
    \label{range_profile_3_targets}
\end{figure}

Notably, $\widetilde{\mathsf{y}}_s(\tau)$ can be interpreted as a linear combination of $Q$ time-shifted versions of $\mathsf{R}(\tau)$, with added noise, often referred to as the \textit{range profile} in radar literature \cite{Richards2005Fundamentals}. To detect the targets, one typically identifies $Q$ peaks in the squared MF output $|\widetilde{\mathsf{y}}_s(\tau)|^2$, where an example is portrayed in Fig. \ref{range_profile_3_targets} for ranging with OFDM signal using 16-PSK constellation and root-raised cosine (RRC) pulse shaping, including $Q = 3$ targets with varying amplitudes located at 10m, 20m, and 25m, respectively. The target detection is achieved using thresholding algorithms such as the constant false-alarm rate (CFAR) detector \cite{Richards2005Fundamentals}, under an SNR = 0 dB. In order to improve the sensing performance, it is desirable for $|\widetilde{\mathsf{y}}_s(\tau)|^2$ to exhibit high peaks at $\tau = \tau_q$ while maintaining low sidelobes elsewhere, which benefits both target detection and estimation.

To further enhance the sensing performance, the coherent integration technique can be employed to effectively reduce both sidelobe and noise levels. In this scheme, the ISAC Tx transmits $M$ i.i.d. information symbol sequences, denoted as $\mathbf{s}^{(0)}, \mathbf{s}^{(1)}, \ldots, \mathbf{s}^{(M-1)}$, by emitting $M$ ISAC signals $\widetilde{\mathsf{x}}^{(0)}(t), \widetilde{\mathsf{x}}^{(1)}(t), \ldots, \widetilde{\mathsf{x}}^{(M-1)}(t)$. Assume that the target parameters $\left\{\gamma_q\right\}_{q=1}^Q$ and $\left\{\tau_{s,q}\right\}_{q=1}^Q$ remain constant over the $M$ transmission slots, such that the sensing Rx can generate $M$ range profiles $\widetilde{\mathsf{y}}_s^{(0)}(\tau), \widetilde{\mathsf{y}}_s^{(1)}(\tau), \ldots, \widetilde{\mathsf{y}}_s^{(M-1)}(\tau)$ through matched-filtering the $M$ received echoes. By coherently integrating these MF outputs, we obtain:
\begin{align}\label{coherent_integration}
    \frac{1}{M}\sum_{m = 0}^{M-1}\widetilde{\mathsf{y}}_s^{(m)}(\tau) = &\nonumber \frac{1}{M}\sum_{q = 1}^Q \gamma_q \sum_{m = 0}^{M-1}\mathsf{R}^{(m)}(\tau-\tau_{s,q})\\
    &+ \frac{1}{M}\sum_{m = 0}^{M-1}\widetilde{\mathsf{z}}_s^{(m)}(\tau),\quad 0\le \tau \le NT.
\end{align}
where $\mathsf{R}^{(m)}(\tau)$ is the ACF of $\widetilde{\mathsf{x}}^{(m)}(t)$, and $\widetilde{\mathsf{z}}_s^{(m)}(\tau)$ represents the output noise from the $m$-th matched filter.


Observe that the noise term in \eqref{coherent_integration} has been averaged, resulting in a reduction of its variance. Furthermore, the overall target sensing performance now critically depends on the geometry of the coherently integrated ACF, defined as: 
\begin{equation}\label{CI_ACF}
    \overline{\mathsf{R}}(\tau): =  \frac{1}{M}\sum_{m = 0}^{M-1}\mathsf{R}^{(m)}(\tau), \quad 0\le \tau \le NT.
\end{equation}
As will be shown later, the sidelobe levels of $\overline{\mathsf{R}}(\tau)$ are also suppressed due to the independence among the $M$ realizations. Due to the randomness of communication symbols carried by the ISAC signal $\widetilde{\mathsf{x}}^{(m)}(t)$, $\overline{\mathsf{R}}(\tau)$ becomes a random function. Therefore, it is more appropriate to analyze its statistical properties rather than focusing on a single realization. In particular, we aim to characterize the expected value of the squared magnitude of $\overline{\mathsf{R}}(\tau)$, given that $\overline{\mathsf{R}}(\tau)$ is a complex-valued function. This expectation is defined as:
\begin{equation}\label{E_squared_ACF}
\mathbb{E}\left(|\overline{\mathsf{R}}(\tau)|^2\right) = \mathbb{E}\left(\left|\frac{1}{M}\sum_{m = 0}^{M-1}\int {\widetilde{\mathsf{x}}^{(m)}(t){\widetilde{\mathsf{x}}^{(m)\ast}}\left( {t - \tau} \right)} dt\right|^2\right).
\end{equation}
In the following subsections, we review recent research efforts in deriving closed-form expressions for \eqref{E_squared_ACF} for a generic communication signal, and provide useful insights for designing ISAC signals based on the mathematical structure of \eqref{E_squared_ACF}.

\subsection{Characterization of the ISAC ACF}
\subsubsection{Discretization of the ACF}
For ease of analysis, let us commence by discretizing the ISAC signal $\widetilde{\mathsf{x}}(t)$ over a temporal grid of $T_s = \frac{T}{L}$, yielding
\begin{equation}\label{baseband_ISAC_signal_discrete}
\widetilde{\mathsf{x}}(kT_s) = \sum\limits_{n = 0}^{N-1}\mathsf{x}_n p(kT_s-nT) = \sum_{n = 0}^{N-1} \mathsf{x}_n \delta(kT_s - nT) \circledast p(kT_s), 
\end{equation} 
where $k = 0,1,\ldots,LN-1$, and $\circledast$ denotes the cyclic convolution due to the addition of a CP. Note here that $L>1$ is required to capture the impact of pulse shaping on sensing performance. If $L = 1$, the discretization simplifies to $\mathbf{x} = \bm{U}\mathbf{s}$ due to the zero-ISI property of the Nyquist pulse. This, however, neglects the contributions of the pulse shaping filter on the mainlobe and sidelobes of the ACF.

Define $p_k := p(kT_s),\;\text{and}~\widetilde{\mathsf{x}}_k :=\widetilde{\mathsf{x}}(kT_s)$, so that the discrete versions of the pulse and baseband signal can be represented as vectors $\bm{p} = \left[p_0,p_1,\ldots,p_{LN-1}\right]^T$ and $\widetilde{\mathbf{x}} = \left[\widetilde{\mathsf{x}}_0,\widetilde{\mathsf{x}}_1,\ldots,\widetilde{\mathsf{x}}_{LN-1}\right]^T$, with the energy of the pulse being normalized to $\left\|\bm{p}\right\|^2 = 1$. In a practical communication Tx, pulse shaping can be implemented digitally through an up-sampling and interpolation procedure applied to $\mathbf{x}$, recasting \eqref{baseband_ISAC_signal_discrete} into a vector formulation as
\begin{equation}
    \widetilde{\mathbf{x}} = \bm{P}\mathbf{x}_{\rm up},
\end{equation}
where 
\begin{align}
\mathbf{x}_{\rm up} = \left[{\mathsf{x}}_0,\bm{0}_{L-1}^T, {\mathsf{x}}_1,\bm{0}_{L-1}^T,\ldots,{\mathsf{x}}_{N-1},\bm{0}_{L-1}^T\right]^T,
\end{align}
and $\bm{P}\in \mathbb{C}^{LN \times LN}$ is a circulant matrix defined as
\begin{equation}
{\bm{P}} = \left[ {\begin{array}{*{20}{c}}
  {{p_0}}&{{p_{LN-1}}}& \ldots &{{p_1}} \\ 
  {{p_1}}&{{p_0}}& \ldots &{{p_2}} \\ 
   \vdots & \vdots & \ddots & \vdots  \\ 
  {{p_{LN-1}}}&{{p_{LN-2}}}& \ldots &{{p_0}} 
\end{array}} \right],
\end{equation}
which interpolates $\mathbf{x}$ to a higher resolution before transmission. Accordingly, the discrete version of the ACF ${\mathsf{R}}(\tau)$ is given by
\begin{equation}
    \mathsf{R}_k = \widetilde{\mathbf{x}}^H\bm{J}_k\widetilde{\mathbf{x}} = \mathbf{x}_{\rm up}^H\bm{P}^H\bm{J}_k\bm{P}\mathbf{x}_{\rm up},\;\; k = 0,1,\ldots,LN-1,
\end{equation}
where $\bm{J}_k\in\mathbb{C}^{LN\times LN}$ is the $k$-th periodic time-shift matrix, defined by \cite{4838816}
\begin{equation}
    \bm{J}_k = \left[ {\begin{array}{*{20}{c}}
  {\mathbf{0}}&{{{\bm{I}}_{LN-k}}} \\ 
  {\bm{I}_{k}}&{\mathbf{0}} 
\end{array}} \right],
\end{equation}
and
\begin{equation}
    \bm{J}_{-k} = \bm{J}_{LN-k} = \bm{J}_k^T= \left[ {\begin{array}{*{20}{c}}
  {\mathbf{0}}&{{{\bm{I}}_{k}}} \\ 
  {\bm{I}_{LN-k}}&{\mathbf{0}} 
\end{array}} \right].
\end{equation}
Note that the periodicity in $\bm{J}_k$ is again due to the addition of the CP. It follows that the discrete version of the coherently integrated ACF \eqref{CI_ACF} becomes
\begin{equation}\label{CI_Rk}
    \overline{\mathsf{R}}_k = \frac{1}{M}\sum_{m = 0}^{M-1}\widetilde{\mathbf{x}}^{(m)H}\bm{J}_k\widetilde{\mathbf{x}}^{(m)}.
\end{equation}

\subsubsection{The ``Iceberg in the Sea'' Structure of the ACF}
To shed light on the communication-centric ISAC transmission under random signaling, the work of \cite{liu2025iceberg} derived a closed form of the expectation of $\left|\overline{\mathsf{R}}_k\right|^2$, given by
\begin{align}\label{CIed_squared_P-ACF}
    &\nonumber\mathbb{E}(|\overline{\mathsf{R}}_k|^2)  = \underbrace {N{{\left| {\widetilde{\bm{f}}_{k + 1}^H\widetilde{\bm{g}}_k} \right|}^2}}_{\text{Iceberg}} \\
    &\nonumber + \underbrace {\frac{1}{M}\left\{\left\| { \widetilde{\bm{g}}_k} \right\|^2 + ({\kappa} - 2)N\left\| {\widetilde {\bm{V}}^T\left( {\widetilde{\bm{g}}_k\odot \widetilde{\bm{f}}_{k + 1}^\ast} \right)} \right\|^2\right\}}_{\text{Sea Level}},\\
    & = \left|\mathbb{E}({\overline{\mathsf{R}}_k})\right|^2 + \operatorname{var}(\overline{\mathsf{R}}_k),\quad k = 0,1,\ldots,LN-1,
\end{align}
where $\widetilde{\bm{f}}_{k + 1} \in \mathbb{C}^{N \times 1}$ contains the first $N$ entries of the $(k+1)$-th column of the size-$LN$ DFT matrix $\bm{F}_{LN}$, and $\widetilde{\bm{V}} \in \mathbb{R}^{N \times N}$ is defined as: 
\begin{equation} 
\widetilde{\bm{V}} = (\bm{F}_N \bm{U}) \odot (\bm{F}_N^\ast \bm{U}^\ast), 
\end{equation} 
where $\bm{U}$ is the modulation basis. Since $\widetilde{\bm{V}}$ is generated by the entry-wise square of an unitary matrix $\bm{F}_N \bm{U}$, it becomes a bi-stochastic matrix with nonnegative real entries \cite{chterental2008orthostochastic}, each of whose rows and columns sums to 1. Moreover, $\kappa$ denotes the \textit{kurtosis} of the adopted constellation $\mathcal{S}$, defined as \cite{liu2024OFDM}:
\begin{equation}
    \kappa := \frac{\mathbb{E}\left\{|\mathsf{s}-\mathbb{E}(\mathsf{s})|^4\right\}}{\mathbb{E}\left\{|\mathsf{s}-\mathbb{E}(\mathsf{s})|^2\right\}^2} =  \mathbb{E}(|\mathsf{s}|^4),\quad \forall~ \mathsf{s}\in\mathcal{S},
\end{equation}
which is the normalized fourth moment of the constellation. Finally, due to the folded spectrum criterion  and that the roll-off factor $\alpha \le 1$, the vector $\widetilde {\bm{g}}_k \in \mathbb{C}^{N \times 1}$ is determined by the squared spectrum of pulse $\bm{p}$ in the following manner:
\begin{equation}\label{gk_def}
    \widetilde{\bm{g}}_k = \bm{g} + (\mathbf{1}_N-\bm{g})e^{-\frac{j2\pi k}{L}},
\end{equation}
where $\bm{g} = \left[g_0,g_1,\ldots,g_{N-1}\right]^T$ contains the first $N$ samples of the squared spectrum $N(\bm{F}_{LN}\bm{p})\odot(\bm{F}_{LN}^\ast\bm{p}^\ast)\in \mathbb{C}^{LN \times 1}$. Notably, the impact of all three signaling blocks—modulation basis, constellation, and pulse shaping—on the shape of the ACF is well-captured in \eqref{CIed_squared_P-ACF}. 

\begin{figure}[!t]
	\centering
	\includegraphics[width = \columnwidth]{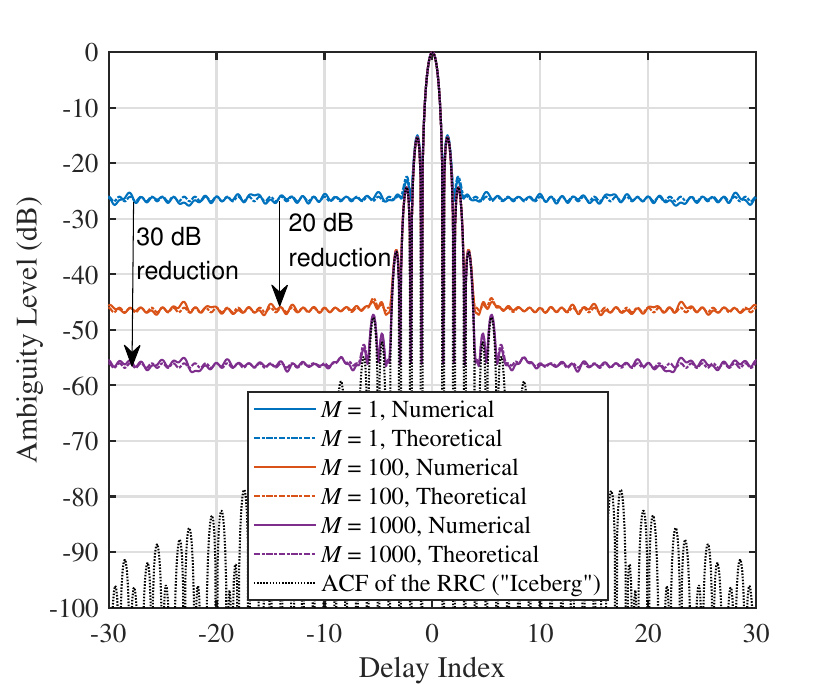}
	\caption{The average squared ACF and corresponding coherent integration versions of an OFDM signal, with 16-QAM constellation and $\alpha = 0.35$ RRC pulse shaping, $N = 128$, $L = 10$, $M = 1, 100, \text{and}\;1000$.}
    \label{iceberg_example}
\end{figure}

The equation \eqref{CIed_squared_P-ACF} reveals an intriguing ``iceberg-in-the-sea'' structure. The ``iceberg'' portion represents the squared mean of $\overline{\mathsf{R}}_k$, which can be rigorously shown to correspond to the squared ACF of the pulse itself \cite{liu2025iceberg}, expressed as: 
\begin{equation}
    N{{| {\widetilde{\bm{f}}_{k + 1}^H\widetilde {\bm{g}}_k} |}^2} = \left|\bm{p}^H\bm{J}_k\bm{p}\right|^2,\quad k = 0,1,\ldots,LN-1.
\end{equation}
This ``iceberg'' component defines the overall shape of \eqref{CIed_squared_P-ACF}. Meanwhile, the ``sea level'' aspect is driven by the variance of $\overline{\mathsf{R}}_k$, which arises largely from the randomness of communication symbols. By coherently integrating $M$ MF outputs derived from $M$ i.i.d. ISAC signals, the ``sea level'' can be significantly reduced by a factor of $M$.

\subsubsection{Example}
Fig. \ref{iceberg_example} demonstrates an example using OFDM signaling with a 16-QAM constellation and $\alpha = 0.35$ RRC pulse shaping, under an $L = 10$ over-sampling ratio. Here, we compare the average squared ACFs for various values of $M$. Notably, the ACF of the random OFDM signal aligns closely with the pulse’s ACF near the mainlobe region, representing the ``tip'' of the ``iceberg''. Beyond this region, the sidelobes are dominated by the ``sea level'' component. By increasing the number of coherent integrations from $M = 1$ to $100$ and then to $1000$, we observe an obvious reduction in the ``sea level'' by 20 dB and 30 dB, respectively, thereby revealing more of the ``iceberg'' component in the average squared ACF.

Building on these observations, we next review recent advances in optimizing modulation basis, constellation design, and pulse shaping for random ISAC signals. These enhancements aim to boost sensing performance while preserving optimal communication quality, or to establish a scalable tradeoff between S\&C.

\section{Waveform Design for Random ISAC Signals}\label{waveform_design_sec}
In this section, we present design guidelines for three core building blocks in communication-centric ISAC systems, focusing on reshaping the statistical properties of the ACF of random ISAC signals. Specifically, our objective is to minimize the average peak-to-sidelobe level ratio (PSLR) of $\overline{\mathsf{R}}_k$ in \eqref{CI_Rk}, thereby enhancing multi-target sensing performance in the range domain within the MF framework. The estimation of other critical parameters, such as Doppler and angle, using random signals is left as a topic for future research.

According to \cite{liu2024OFDM,liu2025iceberg}, the average mainlobe level of the ACF under arbitrary modulation basis, constellation mapping, and pulse shaping filter, can be expressed in closed form as:
\begin{equation}\label{mainlobe_0}
\mathbb{E}(|\overline{\mathsf{R}}_0|^2) = N^2 + \frac{(\kappa - 1)N}{M},
\end{equation}
indicating that the mainlobe level is determined solely by the kurtosis of the constellation. Furthermore, when the coherent integration number $M$ is sufficiently large, the impact of the kurtosis becomes negligible, resulting in an approximately constant average mainlobe level of $N^2$. Recognizing this, it suffices to optimize only the average sidelobe level, which can be formulated as the following generic optimization problem:
\begin{align}\label{generic_waveform_design}
    \mathop {\min }\limits_{\bm{U}\in\mathbb{U}(N),\;P_{\mathsf{s}}(s),\;0 \le {\bm{g}} \le 1} \;\; \mathbb{E}(|\overline{\mathsf{R}}_k|^2), \quad \forall~ k \in \mathcal{K}_{\text{SL}},
\end{align}
where $\bm{U}$ and $\bm{g}$ denote the modulation basis and squared spectrum of the pulse, respectively, as defined above, $P_{\mathsf{s}}(s)$ represents the input constellation distribution, and $\mathcal{K}_{\text{SL}}$ refers to the sidelobe region. While seeking for the globally optimal solution of \eqref{generic_waveform_design} remains a highly challenging task, in what follows, we elaborate on the general design methodology of each component.

\subsection{Modulation Basis Design}\label{modulation_sec}
From \eqref{CIed_squared_P-ACF}, it is evident that the modulation basis influences the sidelobe level solely through the squared norm term $\| {\widetilde {\bm{V}}^T( {\widetilde{\bm{g}}_k\odot \widetilde{\bm{f}}_{k + 1}^\ast})} \|^2$ in the ``sea level'' part. For a given pair of pulse shaping filter and constellation, \eqref{generic_waveform_design} simplifies to:
\begin{align}\label{modulation_basis_design_generic}
    &\nonumber \mathop {\min }\limits_{\bm{U}\in\mathbb{U}(N)} \;\; ({\kappa} - 2)\left\| {\widetilde {\bm{V}}^T\left( {\widetilde{\bm{g}}_k\odot \widetilde{\bm{f}}_{k + 1}^\ast} \right)} \right\|^2\\
    &\;\;\;\text{s.t.}\;\;\;\;\;\;\; \widetilde{\bm{V}} = (\bm{F}_N \bm{U}) \odot (\bm{F}_N^\ast \bm{U}^\ast).
\end{align}
Clearly, the optimal modulation basis depends on the sign of $(\kappa - 2)$, also referred to as the {\textit{excess kurtosis}} \cite{decarlo1997meaning}. Notably, if the constellation follows a standard complex Gaussian distribution, i.e., $\mathsf{s}\sim\mathcal{CN}(0,1)$, the kurtosis equals 2. In such a case, the average sidelobe level at each lag $k$ remains constant irrespective of the chosen modulation basis, as the standard Gaussian distribution is invariant under unitary transformations. Inspired by this, we classify constellations into two categories: sub-Gaussian ($\kappa < 2$) and super-Gaussian ($\kappa > 2$). It is worth noting that commonly used constellations, e.g., QAM and PSK, are sub-Gaussian, with their kurtosis values summarized in Table \ref{tab: kurtosis}. Meanwhile, super-Gaussian constellations may be advantageous in scenarios demanding high energy efficiency or where non-coherent communication methods are employed \cite{923716,1532206,1532207}. Typical examples include index modulation and APSK constellations with exponentially growing radii \cite{liu2024OFDM}.

\begin{table}[!t]
\caption{Kurtosis values of typical sub-Gaussian constellations}
\label{tab: kurtosis}
\begin{tabular}{l|c|c|c|c}
\hline
\textbf{Constellation} & PSK     & 16-QAM  & 64-QAM   & 128-QAM  \\ \hline
\textbf{Kurtosis}      & 1       & 1.32    & 1.381   & 1.3427   \\ \hline
\textbf{Constellation} & 256-QAM & 512-QAM & 1024-QAM & 2048-QAM \\ \hline
\textbf{Kurtosis}      & 1.3953  & 1.3506  & 1.3988   & 1.3525   \\ \hline
\end{tabular}
\end{table}

\begin{figure}[!t]
	\centering
	\includegraphics[width = 0.95\columnwidth]{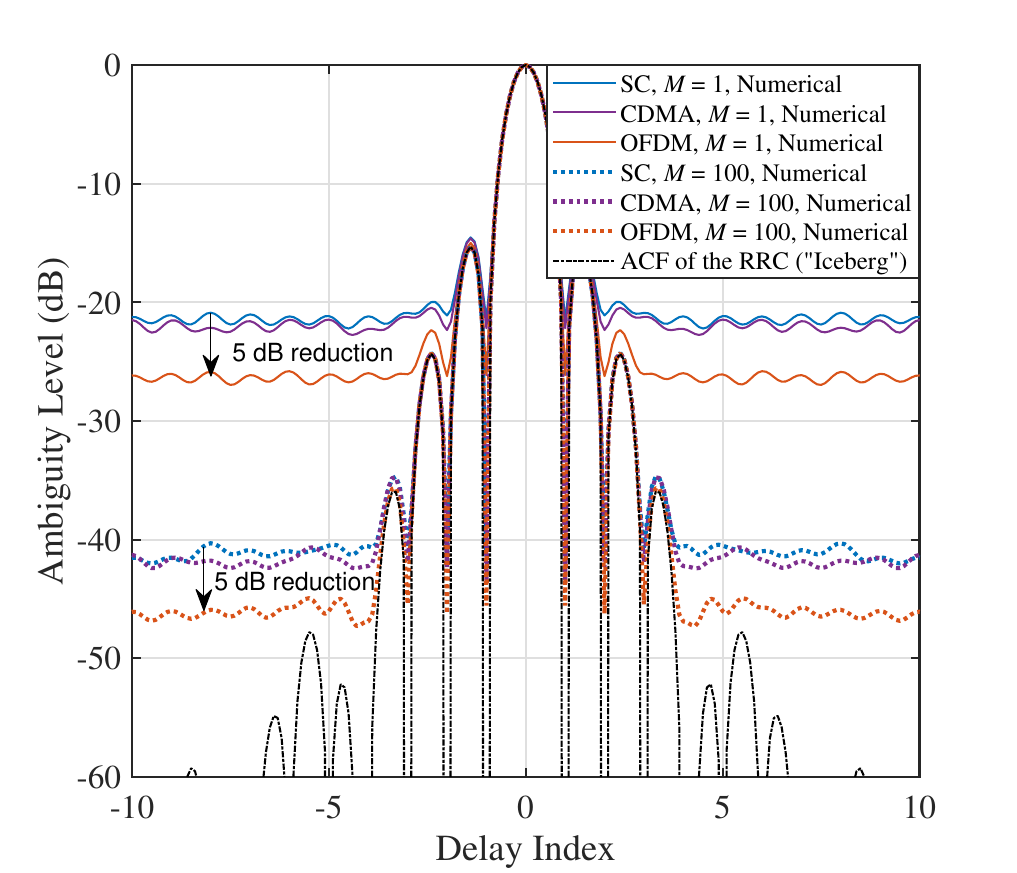}
	\caption{The average squared ACF and corresponding coherent integration versions of SC, CDMA, and OFDM signals, with 16-QAM constellation and $\alpha = 0.35$ RRC pulse shaping, $N = 128$, $L = 10$, $M = 1\; \text{and}\;100$.}
    \label{modulation_example}
\end{figure}

\begin{figure*}[ht!]
    \centering
    \subfloat[]{
    \includegraphics[width=0.26\textwidth]{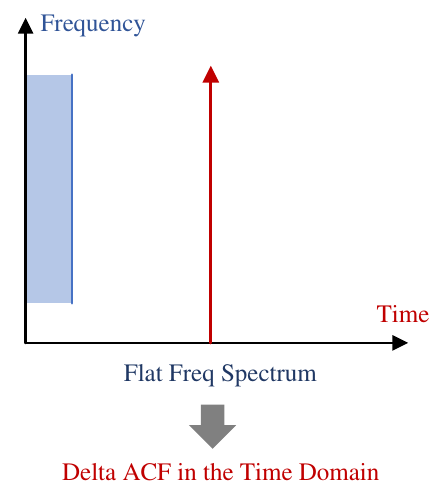} \label{PHY_interpretation_a}
    }
    \subfloat[]{
    \includegraphics[width=0.26\textwidth]{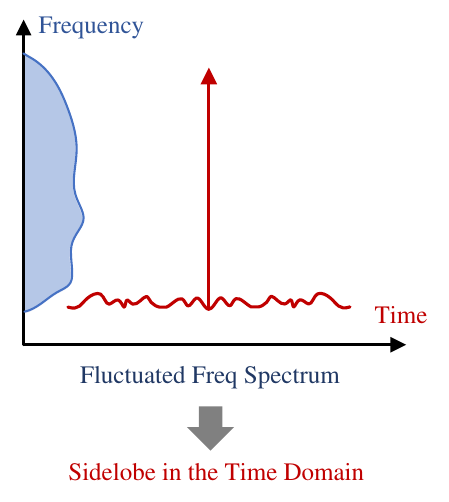}
    \label{PHY_interpretation_b}
    }
    \label{SenOnly_SGP_WF_DDP_SNR}
    \subfloat[]{
    \includegraphics[width=0.26\textwidth]{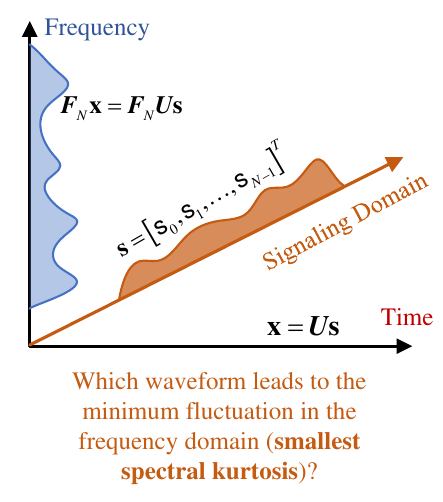}
    \label{PHY_interpretation_c}
    }\caption{Physical interpretation of the optimality of OFDM and SC modulation. (a) The ideal case: A completely flat spectrum leads to a Dirac-Delta ACF; (b) Random data payload causes variability in the squared spectrum, introducing random sidelobes in the ACF; (c) The optimal modulation basis minimizes the fluctuation in the squared spectrum, which is measured by the kurtosis of $\bm{F}_N\bm{U}\mathbf{s}$, where $\bm{F}_N\bm{U}$ is a unitary transform.}
    \label{PHY_interpretation}
\end{figure*} 

\subsubsection{Ranging-Optimal Modulation for Sub-Gaussian Constellations}
Let us first discuss the modulation design for sub-Gaussian constellations. In this case, problem \eqref{modulation_basis_design_generic} reduces to:
\begin{align}\label{modulation_basis_subGaussian}
    &\nonumber \mathop {\max }\limits_{\bm{U}\in\mathbb{U}(N)} \;\; \left\| {\widetilde {\bm{V}}^T\left( {\widetilde{\bm{g}}_k\odot \widetilde{\bm{f}}_{k + 1}^\ast} \right)} \right\|^2\\
    &\;\;\;\text{s.t.}\;\;\;\;\;\;\; \widetilde{\bm{V}} = (\bm{F}_N \bm{U}) \odot (\bm{F}_N^\ast \bm{U}^\ast).
\end{align}
Although problem \eqref{modulation_basis_subGaussian} is generally non-convex, it was proven in \cite{liu2025iceberg} that the globally optimal solution exhibits the following structure:
\begin{equation}\label{U_SUB}
    \bm{U}_{\text{sub}}^\star = \bm{F}_N^H\bm{\varPi}\operatorname{Diag}(\bm{\theta}),
\end{equation}
where $\bm{\varPi}$ is any size-$N$ permutation matrix, and $\bm{\theta}\in\mathbb{C}^{N \times 1}$ can be any vector with unit-modulus entries. This corresponds to OFDM modulation, subject to permutation and phase shifts in the subcarriers. Recalling that $\widetilde {\bm{V}}$ is a bi-stochastic matrix, the optimality of \eqref{U_SUB} is established by showing that the objective function of \eqref{modulation_basis_subGaussian} is maximized when $\widetilde {\bm{V}}$ is any real permutation matrix $\bm{\varPi}$. This result is derived by using the Schur-convexity of the $\ell_2$ norm and the concept of majorization \cite{9256999,9257097,6963438}. More technical details can be found in \cite{liu2025iceberg}.

Overall, the result in \eqref{U_SUB} suggests the following:
\textit{For sub-Gaussian constellations, OFDM achieves the lowest average ranging sidelobe at every lag.}

\subsubsection{Ranging-Optimal Modulation for Super-Gaussian Constellations}
Let us now move to the super-Gaussian constellations with $\kappa > 2$. In such a case, problem \eqref{modulation_basis_design_generic} becomes
\begin{align}\label{modulation_basis_superGaussian}
    &\nonumber \mathop {\min }\limits_{\bm{U}\in\mathbb{U}(N)} \;\; \left\| {\widetilde {\bm{V}}^T\left( {\widetilde{\bm{g}}_k\odot \widetilde{\bm{f}}_{k + 1}^\ast} \right)} \right\|^2\\
    &\;\;\;\text{s.t.}\;\;\;\;\;\;\; \widetilde{\bm{V}} = (\bm{F}_N \bm{U}) \odot (\bm{F}_N^\ast \bm{U}^\ast).
\end{align}
By exploiting again the Schur-convexity of the $\ell_2$ norm, and applying a similar argument, one may readily show that the objective is minimized if the bi-stochastic matrix is uniform, namely, $\widetilde {\bm{V}} = \frac{1}{N}\mathbf{1}\mathbf{1}^T$. Accordingly, the optimal modulation basis takes the form of 
\begin{equation}\label{U_SUPER}
    \bm{U}_{\text{super}}^\star = \bm{\varPi}\operatorname{Diag}(\bm{\theta}),
\end{equation}
corresponding to an SC modulation subject to arbitrary permutation and phase shifts of time-domain symbols. 

Overall, the result in \eqref{U_SUPER} suggests that: \textit{For super-Gaussian constellations, SC achieves the lowest average ranging sidelobe at every lag}.

\subsubsection{Example}
We present an example in Fig. \ref{modulation_example} to validate the optimality of OFDM for sub-Gaussian constellations, using the standard 16-QAM alphabet with a kurtosis of 1.32. The average squared ACF is compared across three modulation bases: OFDM, SC, and CDMA2000 (where $\bm{U}$ is a Hadamard matrix), transmitting $N = 128$ i.i.d. symbols with an $L = 10$ over-sampling ratio, and employing an RRC pulse shaping filter with $\alpha = 0.35$. As predicted by the theoretical results, Fig. \ref{modulation_example} demonstrates that OFDM produces the lowest sidelobe at every lag, achieving a 5 dB improvement over both SC and CDMA. Additionally, after coherently integrating i.i.d. MF outputs for $M = 100$ times, a 20 dB reduction in the sidelobe level is observed in the ``sea level'' region for all signaling schemes.


\subsubsection{Discussion on Optimality of OFDM/SC}
The aforementioned results provide valuable design insights for the modulation formats of communication-centric ISAC systems. While these findings are mathematically rigorous, a further understanding of the optimality of OFDM/SC requires addressing the following critical question: What is the \textit{physical interpretation} underlying these mathematical results? 

To depict the physical insight, we examine the problem through the lens of the Fourier duality and central limit theorem (CLT). Consider a signal with infinite bandwidth, as illustrated in Fig. \ref{PHY_interpretation_a}, which exhibits a perfectly flat amplitude spectrum. According to the Fourier duality, this implies that its ACF is a Dirac-Delta function in the delay domain. This represents the ideal sensing signal for ranging, as it leads to a perfect MF output with no ambiguity. However, as shown in Fig. \ref{PHY_interpretation_b}, the presence of random fluctuations in the communication data payload causes variability in the squared spectrum of the ISAC signal, which, in turn, causes random sidelobes in the ACF of Fig. \ref{PHY_interpretation_b}. Intuitively, this suggests that minimizing the average ranging sidelobe level of an ISAC signal is equivalent to minimizing the fluctuations in its frequency-domain representation, which can be quantified by the variance of the squared spectrum, and is proportional to the frequency-domain kurtosis of the signal. Thus, reducing ranging sidelobes can be achieved by minimizing the signal’s frequency-domain kurtosis.

To explore this further, we re-examine the ISAC modulation basis design as presented in Fig. \ref{PHY_interpretation_c}. As outlined in the generic model in \eqref{time_domain_discrete_sample}, a modulation basis can be interpreted as a unitary rotation $\bm{U}$ applied to the i.i.d. symbol sequence $\mathbf{s}$. Accordingly, the corresponding frequency-domain digital samples are given by $\bm{F}_N\bm{U}\mathbf{s}$, where the product $\bm{F}_N\bm{U}$ remains unitary. Based on the CLT, a linear transform applied to a random vector with i.i.d. entries results in a distribution that asymptotically approaches a Gaussian form. As a consequence, any unitary transform $\bm{F}_N\bm{U}$ increases the kurtosis of i.i.d. sub-Gaussian symbols (e.g., QAM and PSK), while it decreases the kurtosis of i.i.d. super-Gaussian symbols. Therefore, to minimize the kurtosis of the vector $\bm{F}_N\bm{U}\mathbf{s}$, where $\mathbf{s}$ is sub-Gaussian, the optimal strategy is to retain its kurtosis, which is achieved by setting $\bm{F}_N\bm{U} = \bm{I}_N$, corresponding to the OFDM modulation. In contrast, if $\mathbf{s}$ is super-Gaussian, since any unitary transform would reduce its kurtosis, the optimal strategy is to maximize the rotation over $\mathbf{s}$. In this case, the appropriate transformation is $\bm{F}_N \bm{U} = \bm{F}_N$, leading to the SC modulation.

\subsection{Constellation Design}\label{PCS_sec}
Now, we shift our focus from modulation format to constellation design. As noted in \eqref{CIed_squared_P-ACF}, $\mathbb{E}(|\overline{\mathsf{R}}_k|^2)$ depends on the constellation solely through its kurtosis. For a given pair of modulation format and pulse shaping filter, the minimum kurtosis (and consequently sidelobes) can be achieved with any PSK constellation, given the fact that $\kappa \ge 1$. However, PSK may result in lower communication rates compared to its QAM counterpart of the same order, highlighting again the DRT in ISAC systems \cite{Xiong_TIT}.

To balance the achievable rate for communication and ranging sidelobe level for sensing, a practical approach is \textit{constellation shaping} \cite{9460990,8640810}, which is originally tailored for improving the spectral and energy efficiencies of digital communication systems. Constellation shaping techniques can be broadly classified into two categories: probabilistic constellation shaping (PCS) and geometric constellation shaping (GCS). PCS modifies the constellation's probability density function (PDF) through a distribution matcher, which maps the bit stream to the desired probability distribution of the constellation. In contrast, GCS directly optimizes the amplitudes and phases of constellation symbols themselves. Both techniques are capable of reshaping the statistical characteristics of the adopted constellation.

\subsubsection{Probabilistic Constellation Shaping for ISAC}
For the ISAC constellation design problem, we focus on the PCS approach, which can be formulated as an optimization problem aimed at maximizing the communication MI under a ranging sidelobe level threshold. This can be expressed as \cite{10685511}:
\begin{align}\label{MI_PCS_design}
    &\nonumber \mathop {\max }\limits_{P_{\mathsf{s}}(s)} \;\; I(\widetilde{\mathbf{y}}_c;\mathbf{s})\\
    &\nonumber\;\text{s.t.}\;\;\;\;\;\mathbb{E}(|\mathsf{s}|^4) \le c_0,\;\; \mathbb{E}(|\mathsf{s}|^2) = 1,\\
    &\;\;\;\;\;\;\;\;\;\; \mathbb{E}(\mathsf{s}) = 0, \;\;\mathbb{E}(\mathsf{s}^2) = 0,\;\;\mathsf{s}\in\mathcal{S},
\end{align}
where $P_{\mathsf{s}}(s)$ stands for the distribution of constellation, ${\mathbf{y}}_c$ represents the MF output signal at the communication receiver, and $c_0 \ge 1$ is a pre-determined constant that controls the kurtosis of the constellation, ensuring that the average ranging sidelobes remain within acceptable bounds. The alphabet $\mathcal{S}$ denotes the set of discrete constellation points, which must satisfy normalized power and rotational symmetry constraints, as outlined in \eqref{regularized_constraints}. 

Problem \eqref{MI_PCS_design} is inherently a functional optimization problem, as the optimization variable $P_{\mathsf{s}}(s)$ is a function defined over $\mathcal{S}$. Indeed, problem \eqref{MI_PCS_design} may be viewed as a specific example of the C-D tradeoff problem in \eqref{CD_IT_meaning}, where the kurtosis constraint acts as a sensing cost function. However, the MI in the objective function does not have a closed-form expression due to the discrete alphabet $\mathcal{S}$. To address this challenge, an optimization-based PCS method was introduced in \cite{10685511} for $M_{\mathsf{s}}$-ary QAM constellations under OFDM modulation. In this case, the multi-path communication channel is diagonalized into $N$ parallel orthogonal AWGN channels, allowing us to focus on the MI of each single scalar AWGN channel, which is denoted as $I(\mathsf{y}_c; \mathsf{s})$.

Let $\bm{p}_{\mathsf{s}} = \left[p_{\mathsf{s},1},p_{\mathsf{s},2},\ldots,p_{\mathsf{s},M_{\mathsf{s}}}\right]^T$ be the probability distribution vector of the considered $M_{\mathsf{s}}$-ary QAM constellation, with $s_m$ being the $m$-th QAM symbol. The MI, by its definition, can be represented as \cite{Yeung2008Information}:
\begin{align}
    &\nonumber I(\mathsf{y}_c;\mathsf{s})  = \sum\limits_{m=1}^{M_{\mathsf{s}}}p_{\mathsf{s},m}\int  p(y_c|s_m)\log\frac{p(y_c|s_m)}{p(y_c)}dy_c\\
    & = \mathop {\max }\limits_{q(s_m|y_c)} \underbrace {\sum\limits_{m=1}^{M_{\mathsf{s}}}p_{\mathsf{s},m}\int  p(y_c|s_m)\log\frac{q(s_m|y_c)}{p(y_c)}dy_c}_{{F(\bm{p}_{\mathsf{s}},\bm{q}_{\mathsf{s}|\mathsf{y}_c})}},
\end{align}
where $q(s_m|y_c)$ is the probability transition function from the received signal set $\mathcal{Y}c$ to the constellation alphabet $\mathcal{S}$, and $\bm{q}_{\mathsf{s}|\mathsf{y}c}$ is its discretized form. The optimization problem \eqref{MI_PCS_design} can thus be reformulated as \cite{10685511}:
\begin{align}\label{PCS_design} 
    &\nonumber \mathop {\max }\limits_{\bm{p}_{\mathsf{s}}}\mathop {\max }\limits_{\bm{q}_{\mathsf{s}|\mathsf{y}_c}} \;\; {F(\bm{p}_{\mathsf{s}},\bm{q}_{\mathsf{s}|\mathsf{y}_c})}\\
    &\nonumber\;\;\text{s.t.}\;\sum\limits_{m=1}^{M_{\mathsf{s}}}p_{\mathsf{s},m}|s_m|^4 \le c_0,\;\; \sum\limits_{m=1}^{M_{\mathsf{s}}}p_{\mathsf{s},m}|s_m|^2 = 1,\\
    &\nonumber\;\;\;\;\;\;\; \sum\limits_{m=1}^{M_{\mathsf{s}}}p_{\mathsf{s},m}s_m = 0, \;\;\sum\limits_{m=1}^{M_{\mathsf{s}}}p_{\mathsf{s},m}s_m^2 = 0,\\
    &\;\;\;\;\;\;\;\sum\limits_{m=1}^{M_{\mathsf{s}}}p_{\mathsf{s},m} = 1, \quad p_{\mathsf{s},m} \ge 0,\quad\forall~ m,
\end{align}
where the last two constraints are enforced since $\bm{p}_{\mathsf{s}}$ is a point on the probability simplex. Here, $F(\bm{p}_{\mathsf{s}}, \bm{q}_{\mathsf{s}|\mathsf{y}c})$ is jointly concave in $\bm{p}_{\mathsf{s}}$ and $\bm{q}_{\mathsf{s}|\mathsf{y}c}$, and all constraints are linear in the probability vector $\bm{p}_{\mathsf{s}}$, making it a convex program. By exploiting this fact, a modified Blahut-Arimoto (MBA) algorithm was proposed in \cite{10685511}. Through constructing the Lagrange multiplier of \eqref{PCS_design}, and iteratively solving for $\bm{p}_{\mathsf{s}}$ and $\bm{q}_{\mathsf{s}|\mathsf{y}_c}$ in an alternative manner, the MBA method ensures efficient convergence to the global optimum.



\subsubsection{Example}
\begin{figure}[!t]
\centering
\subfloat[PCS results for 16-QAM and 64-QAM.]{\includegraphics[width=0.95\columnwidth]{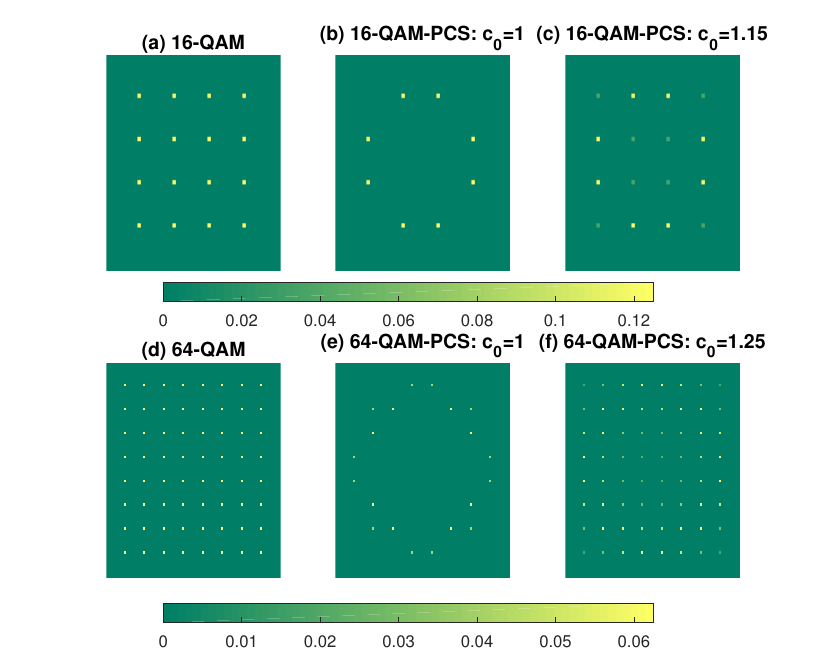}
\label{PCS_results}}
\vspace{0.1in}
\subfloat[An explicit performance tradeoff between S\&C.]{\includegraphics[width=0.95\columnwidth]{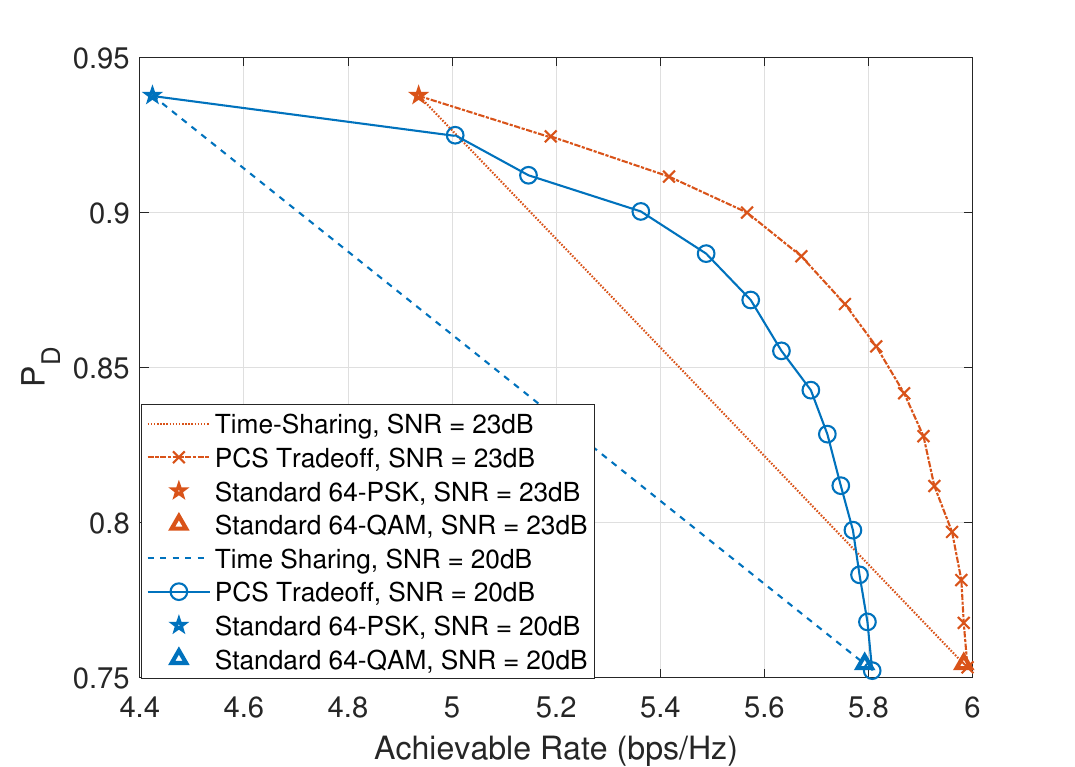}
\label{PCS_results_tradeoff}}
\caption{An illustrative example of the PCS technique for random ISAC signals. (a) PCS results for 16-QAM and 64-QAM under varying values of the kurtosis threshold $c_0$; (b) An explicit performance tradeoff between target detection probability for sensing and achievable rate for communication implemented by adjusting $c_0$ from 1 to 1.381 (kurtosis of the uniform 64-QAM) in the PCS approach, under OFDM modulation and 64-QAM constellation, with different values of transmit SNR.}
\label{PCS_Example}
\end{figure}

We present an illustrative example to demonstrate the effectiveness of the PCS approach for ISAC. Fig. \ref{PCS_results} shows the optimal PCS results for 16-QAM and 64-QAM constellations at different kurtosis thresholds $c_0$, with the probability of each point represented by color depth. As the kurtosis threshold decreases, symbols with nearly or exactly unit modulus are transmitted with higher probabilities, while those on larger or smaller circles are less likely to be used. This aligns with the DRT theory, which suggests that sensing favors constellations with constant modulus. Inevitably, this reduces the communication MI and therefore introduces a graceful tradeoff with the communication rates. Note that reducing the ACF sidelobe level would enhance the weak target detection performance in the range domain, as the sidelobe of strong clutter can significantly interfere with or even mask the mainlobe of weak targets. By realizing this, Fig. \ref{PCS_results_tradeoff} illustrates the explicit S\&C performance of 64-QAM under OFDM modulation across different transmit SNR values, highlighting the achievable communication rate and the detection probability for sensing a weak target in the presence of strong clutter. By adjusting $c_0$, the PCS method provides a scalable tradeoff between S\&C performance metrics, significantly outperforming the naive time-sharing approach between standard uniformly distributed 64-QAM and 64-PSK constellations.

\subsection{Pulse Shaping Design}
We conclude this section by discussing pulse shaping design methodologies for ISAC, given a specific pair of constellation and modulation bases. A closer examination of \eqref{CIed_squared_P-ACF} reveals that $\mathbb{E}(|\overline{\mathsf{R}}_k|^2)$ is a convex quadratic function of $\widetilde{\bm{g}}_k$, and thus convex in the squared spectrum of the pulse, namely, the vector $\bm{g}$. Therefore, one may minimize the sidelobe level within the region $\mathcal{K}_{\text{SL}}$ over the feasible set of $\bm{g}$.

\subsubsection{Generic Pulse Shaping Design}
To proceed, we first discuss the constraints on $\bm{g}$. It is evident that the folded spectrum criterion \eqref{folded_spectrum} is implicitly satisfied in \eqref{gk_def}, ensuring the Nyquist property of the pulse and consequently eliminating the ISI. This guarantees that the communication performance remains unaffected. Moreover, the pulse has a roll-off factor $\alpha \in [0, 1]$, implying that $(1 - \alpha)N$ entries of $\bm{g}$ are either $0$ or $1$. Upon letting $N_{\alpha} = \alpha N$, and assuming that $N - N_{\alpha}$ is even, these constraints can be expressed as
\begin{equation}\label{roll_off_constraint}
g_n = \left\{ \begin{gathered}
  0,\quad \text{if}~0 \le n \le \frac{N - N_{\alpha}}{2} - 1, \hfill \\
  1,\quad \text{if}~\frac{N + N_{\alpha}}{2}\le n \le N-1. \hfill \\ 
\end{gathered}  \right.
\end{equation}
Additionally, to ensure the roll-off part is monotonically increasing, we impose the following constraints:
\begin{equation}\label{monotonic_constraint}
    g_{n+1} - g_{n} \ge 0, \quad \frac{N - N_{\alpha}}{2} \le n \le \frac{N + N_{\alpha}}{2} - 2.
\end{equation}
Finally, the power of the pulse has been normalized, yielding the constraint:
\begin{equation}\label{power_constraint}
    \sum\limits_{n = 0}^{N-1} g_n = \frac{N}{2}.
\end{equation}
Therefore, the generic pulse shaping problem may be formulated as \cite{Liao_Pulse_Shaping}:
\begin{align}\label{generic_pulse_design}
    &\nonumber \mathop {\min }\limits_{0 \le {\bm{g}} \le 1} \;\; \mathbb{E}(|\overline{\mathsf{R}}_k|^2), \quad \forall~ k \in \mathcal{K}_{\text{SL}}\\
    &\;\;\text{s.t.}\;\;\;\;\;\;\eqref{roll_off_constraint}-\eqref{power_constraint},
\end{align}
which is a linearly constrained convex Pareto problem. 

\subsubsection{Iceberg Shaping}
To further simplify the problem, note that the geometry of $\mathbb{E}(|\overline{\mathsf{R}}_k|^2)$ is primarily determined by the ``iceberg'' part when the coherent integration number $M$ is sufficiently large. Based on this observation, one can focus on shaping the ``iceberg'', i.e., the squared ACF of the pulse shaping filter itself, rather than minimizing the sidelobes of both the ``iceberg'' and ``sea level'' components. In this case, the objective is to minimize either the sum of the sidelobes over the region $\mathcal{K}_{\text{SL}}$ of the iceberg, or the maximum sidelobe within this region, yielding the following problem \cite{liu2025iceberg}: 
\begin{align}\label{iceberg_shaping}
    &\nonumber \mathop {\min }\limits_{0 \le {\bm{g}} \le 1} \;\sum_{k\in\mathcal{K}_{\text{SL}}}{{\left| {\widetilde{\bm{f}}_{k + 1}^H\widetilde{\bm{g}}_k} \right|}^2}\;\;\text{or}\;\; \mathop {\max }\limits_{k}{{\left| {\widetilde{\bm{f}}_{k + 1}^H\widetilde{\bm{g}}_k} \right|}^2}\\
    &\;\;\text{s.t.}\;\;\;\;\;\;\eqref{roll_off_constraint}-\eqref{power_constraint},
\end{align}
which is a linearly constrained convex quadratic program that can be efficiently solved via off-the-shelf numerical tools.

\begin{figure}[!t]
\centering
\subfloat[Range estimation performance with/without coherent integration.]{\includegraphics[width=\columnwidth]{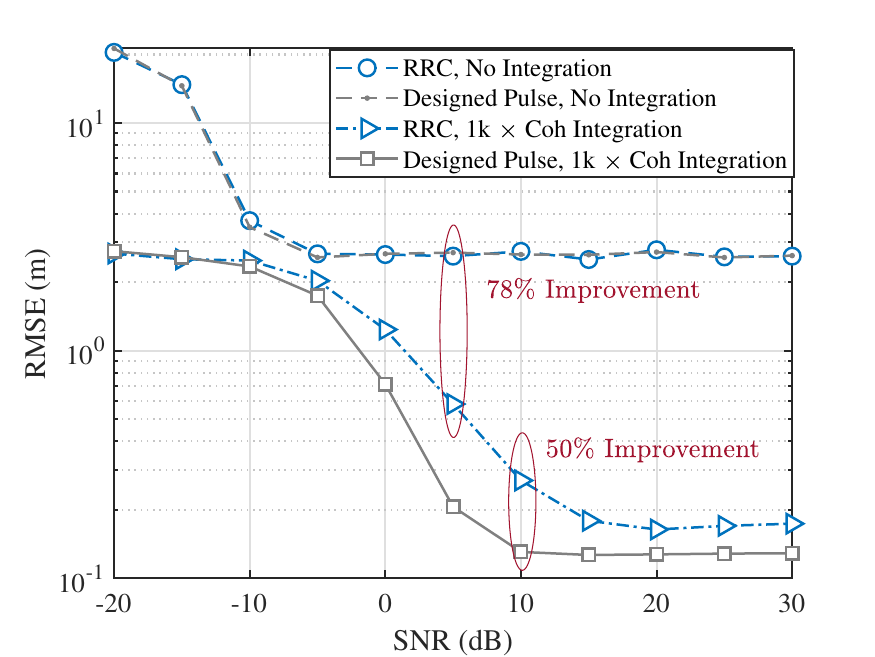}
\label{QAM_vs_NcohIntegration}}
\vspace{0.1in}
\subfloat[Range profiles with/without coherent integration.]{\includegraphics[width=0.9\columnwidth]{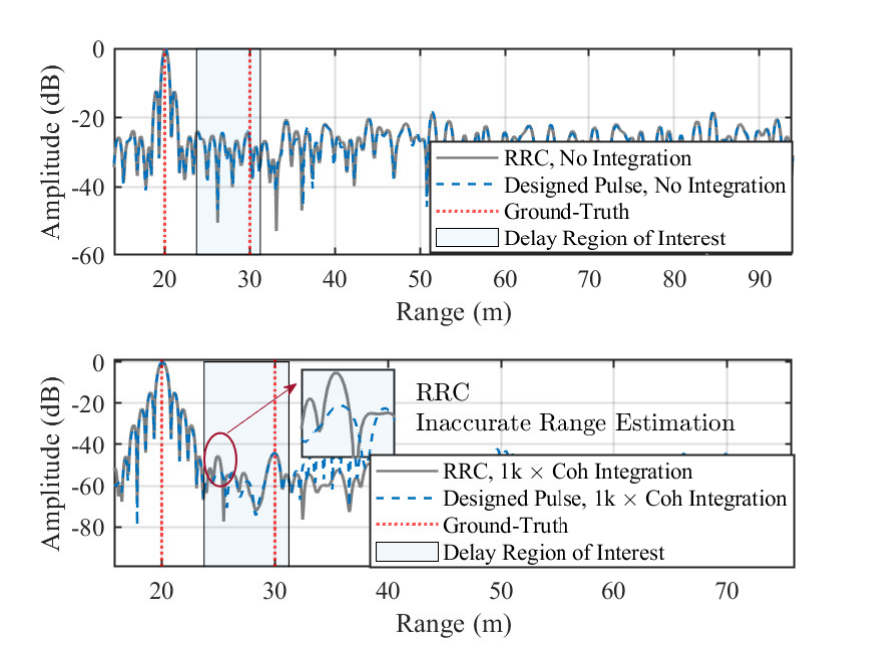}
\label{QAM_RangeProfile}}
\caption{The range estimation performance and profiles of two targets under OFDM with 16-QAM constellation, where $N = 128$, $\alpha = 0.35$, $L = 10$, $M = 1000$, and range region of interest is $\left[23.74\text{m},31.24\text{m}\right]$.}
\label{PSK_QAM_RangeEstimation}
\end{figure}

\subsubsection{Example}
We present a design example of the ISAC pulse shaping filter with coherent integration in Fig. \ref{PSK_QAM_RangeEstimation}, where we apply the iceberg shaping technique in \eqref{iceberg_shaping} to minimize the summation of the ranging sidelobes within the region $\left[23.74\text{m}, 31.24\text{m}\right]$. The ISAC signal adopts OFDM modulation, carrying $N = 128$ i.i.d. 16-QAM symbols, under an over-sampling ratio of $L = 10$. To demonstrate the performance improvement gained from sidelobe reduction in the ``iceberg'', we consider a two-target detection scenario, where one target is located at $20$m and the other at $30$m. The target at $20$m has an amplitude $43 \sim 46$ dB higher than the one at $30$m. The benchmark is the RRC pulse shaping, with a roll-off factor of $\alpha = 0.35$ set for both the RRC and iceberg shaping techniques.

Fig. \ref{QAM_vs_NcohIntegration} shows the ranging root mean squared error (RMSE) curves with and without coherent integration. It can be observed that, before the coherent integration operation, both the RRC and iceberg shaping techniques suffer from poor ranging performance. However, after coherently integrating over $M = 1000$ i.i.d. MF output signals, the ranging RMSE is reduced by more than $78\%$. Additionally, the iceberg shaping method achieves an extra $50\%$ improvement in ranging accuracy compared to its RRC counterpart. This performance improvement is also clearly visible in the corresponding range profiles shown in Fig. \ref{QAM_RangeProfile}. Without coherent integration, the weak target is obscured by the sidelobes generated by the ``sea level'' for both pulse shaping filters, resulting in large range estimation errors. With $1000$ times of coherent integration, the ``sea level'' sidelobes are effectively reduced by $30$ dB, after which the ranging performance is primarily determined by the sidelobes generated from pulses themselves. In this case, the peak corresponding to the weak target can be accurately detected for the ISAC signaling scheme with the designed pulse, thanks to the minimization of the sidelobes within the range region of interest. However, for the RRC pulse shaping, the weak target may be inaccurately located due to the high sidelobe at $24.5$m.

\section{MIMO Precoding with Random ISAC Signals}\label{MIMO_sec}
In this section, we extend our analysis from the previous single-antenna systems to their multi-antenna counterparts, through conceiving dedicated precoders for MIMO ISAC systems under random signaling. Specifically, we introduce the signal model in Sec. \ref{subsection_precoding_signal}, followed by algorithms for solving the precoding design problems in sensing-only and ISAC systems, presented in Sec. \ref{subsection_precoding_sensing} and \ref{subsection_precoding_isac}, respectively.

\subsection{Signal Model and Ergodic LMMSE}\label{subsection_precoding_signal}



Consider a P2P monostatic MIMO ISAC system, with a BS equipped with $N_t$ transmit antennas and $N_s$ receive antennas at its sensing Rx, serving a communication user (CU) with $N_c$ receive antennas while simultaneously detecting one or multiple targets. Assume that target sensing is conducted over a coherent processing interval consisting of $N$ time-domain snapshots. Following the vector Gaussian model in \eqref{vector_Gaussian_model}, the MIMO signal models for S\&C are expressed as  
\begin{subequations}\label{linear_model}
	\begin{align}
	\mathbf{Y}_c = \mathbf{H}_c \mathbf{X}+\mathbf{Z}_c, \label{linear_comm_model} \\
	\mathbf{Y}_s =\mathbf{H}_s \mathbf{X}+\mathbf{Z}_s.         \label{linear_sens_model}
	\end{align}
\end{subequations}  
In the above, $\mathbf{Y}_c \in \mathbb{C}^{N_c \times N}$ represents the received signal matrix at the CU receiver and $\mathbf{Y}_s \in \mathbb{C}^{N_s \times N}$ denotes the echoes at the BS sensing receiver, the matrix $\mathbf{H}_c \in \mathbb{C}^{N_c \times N_t}$ is the P2P MIMO channel and $\mathbf{H}_s \in \mathbb{C}^{N_s \times N_t}$ is the spatial-domain target impulse response (TIR) matrix to be estimated, the matrices $\mathbf{Z}_c \in \mathbb{C}^{N_c \times N}$ and $\mathbf{Z}_s \in \mathbb{C}^{N_s \times N}$ represent additive noise, with each entry following $\mathcal{CN}(0,\sigma_c^2)$ and $\mathcal{CN}(0,\sigma_s^2)$, and $\mathbf{X} \in \mathbb{C}^{N_t \times N}$ denotes the ISAC signal matrix. Additionally, we assume that the TIR matrix $\mathbf{H}_s$ follows a wide-sense stationary random process, such that its statistical correlation matrix $\bm{R}_H = \mathbb{E}\{\mathbf{H}_s^H \mathbf{H}_s\}$ keeps unchanged. 

We now make some remarks on the signal model in \eqref{linear_model}. First, the model in \eqref{linear_model} aligns with the one in \eqref{signal_convolution} in a MIMO-OFDM setting, which can be interpreted as narrowband S\&C signals over a sub-channel. Accordingly, $\mathbf{H}_s$ and $\mathbf{H}_c$ are S\&C channel matrices defined for each sub-carrier. Second, this model may be seen as a special case of the generic vector Gaussian model in \eqref{vector_Gaussian_model}, with the parameter to be estimated being the sensing channel matrix itself, namely, ${\bm \upeta} = \operatorname{vec}\left(\mathbf{H}_s\right)$. For sensing purposes, this section focuses solely on estimating the TIR matrix $\mathbf{H}_s$ for each sub-carrier, which will then be collected across all sub-carriers for further processing to extract delay and angle parameters of the targets. 

Let us provide more elaboration on the ISAC signal matrix $\mathbf{X}$ in \eqref{linear_model}, which is expressed as
\begin{align}\label{XWS}
\mathbf{X} = \bm{W} \mathbf{S},
\end{align}
where $\bm{W}\in \mathbb{C}^{N_t \times N_t}$ is the precoding matrix to be optimized and $\mathbf{S} \in \mathbb{C}^{N_t \times N}$ represents the data payload matrix. Let $f(\bm{W} ; \mathbf{S})$ denote a generic sensing cost function, as described in \eqref{general_cost}. The objective of this section is to design the precoding matrix $\bm{W}$ to optimize the sensing cost function $f(\bm{W} ; \mathbf{S}),$ given the (statistical) information of $\mathbf{S}.$

Analyzing and exploring the structure of the cost function $f(\bm{W} ; \mathbf{S})$ often provides valuable insights into the solution of the corresponding optimization problem \cite{boyd2004convex,10636212}. To facilitate the discussion, we use the linear minimum mean squared error (LMMSE) precoder as a specific example of the sensing cost function $f(\bm{W} ; \mathbf{S})$ in this section. Given the ISAC signal $\mathbf{X}$ and the correlation matrix $\bm{R}_H$ for the sensing channel, the celebrated LMMSE estimator of $\mathbf{H}_s$ is given by:
\begin{align}\label{Estimator_LMMSE}
\hat{\mathbf{H}}_s=\mathbf{Y}_s\left(\mathbf{X}^H \bm{R}_H\mathbf{X}+\sigma_s^2 N_s \bm{I}_{N}\right)^{-1} \mathbf{X}^H \bm{R}_H,
\end{align} 
which results in an MSE expressed as \cite{biguesh2006training}
\begin{align}\label{Cond_MMSE}
 {f}(\bm{W} ; \mathbf{S}) &\overset{\quad}{=} \mathrm{Tr}\left\{\Big(\bm{R}_H^{-1} + \frac{1}{\sigma_s^2 N_s}\bm{W}\mathbf{S}\mathbf{S}^H\bm{W}^H\Big)^{-1}\right\}.
\end{align}
Based on this, the precoding design problem can be straightforwardly formulated as:
\begin{equation}\label{problem:precoding}
\begin{aligned}
\min_{\bm{W}}~& {f}(\bm{W} ; \mathbf{S})\\
\text{\normalfont s.t. }~&\|\bm{W}\|_F^2 \leq P_T,
\end{aligned}
\end{equation}
where $P_T$ is the transmit power budget.

In traditional MIMO radar systems, $\mathbf{S}$ in \eqref{XWS} is a deterministic orthogonal training signal satisfying $\frac{1}{N}\mathbf{S}\mathbf{S}^H = \bm{I}_{N_t}$ \cite{4350230}. In this case, problem \eqref{problem:precoding} simplifies into:
\begin{subequations}
\begin{align}
\min_{\bm{W}}~& \mathrm{Tr}\left\{\Big(\bm{R}_H^{-1} + \frac{1}{\sigma_s^2 N_s}\bm{W}\bm{W}^H\Big)^{-1}\right\}\label{deterministic_LMMSE}\\
\text{\normalfont s.t. }~&\|\bm{W}\|_F^2 \leq P_T.
\end{align}
\end{subequations}
The above problem has a closed-form water-filling solution given by \cite{biguesh2006training}:
\begin{align}\label{LMMSE_Determinstic_Opt_W}
\bm{W}_{\mathsf{WF}} = \sqrt{\frac{\sigma_s^2{N_s}}{N}} \bm{Q}\left\{ \max\left(\mu_0 \bm{I}_{N_t} - \bm{\varLambda}^{-1},\bm{0}\right) \right\}^{\frac{1}{2}},
\end{align}
where $\bm{Q}\bm{\varLambda}\bm{Q}^H$ is the eigenvalue decomposition of $\bm{R}$ and $\mu_0$ is a constant (i.e., the ``water level'') such that $\|\bm{W}_{\mathsf{WF}}\|_F^2 = P_T$.


In sharp contrast to traditional radar systems, ISAC systems must employ random signals for target sensing. Since Gaussian signals achieve the capacity of P2P Gaussian channels as shown in \eqref{linear_comm_model}, we consider Gaussian signaling for ISAC systems as an example. Let $\mathbf{S} = [\mathbf{s}_1, \mathbf{s}_2,\dots, \mathbf{s}_N] \in \mathbb{C}^{N_t \times N}$ denote the transmitted random ISAC signal, where each column is i.i.d. and follows the complex Gaussian distribution with zero mean and covariance $\bm{I}_{N_t}$, i.e., $\mathbf{s}_{\ell} \sim \mathcal{CN}(\mathbf{0}, \bm{I}_{N_t})$. In this case, the objective function in problem \eqref{problem:precoding} becomes a random variable (as it depends on the random variable $\mathbf{S}$). Therefore, it is natural to consider an ergodic LMMSE (ELMMSE) that accounts for the signal randomness, defined as \cite{10596930}:
\begin{align}\label{ELMMSE}
{f}_{\mathsf{ELMMSE}}(\bm{W}): = \mathbb{E}\left\{\mathrm{Tr}\Big[\Big(\bm{R}_H^{-1} +\frac{1}{\sigma_s^2 N_s} \bm{W}\mathbf{S}\mathbf{S}^H\bm{W}^H\Big)^{-1}\Big] \right\},
\end{align} 
where the expectation is performed over $\mathbf{S}$. The ELMMSE may be interpreted as a sensing analogy to the ergodic communication rate \cite{tse2005fundamentals}, which can be regarded as a time average of the MSE achieved by random ISAC signals.

\subsection{Sensing-Only Precoding Design }\label{subsection_precoding_sensing}
In this part, we explore the sensing-only precoding designs under random signaling, which serves as a performance benchmark for the ISAC precoding that will be detailed later on. The corresponding optimization problem may be formulated as: 
\begin{equation}\label{problem:precoding:elmmse}
\begin{aligned}
\min_{\bm{W}}~& {f}_{\mathsf{ELMMSE}}(\bm{W})\\
\text{\normalfont s.t. }~&\|\bm{W}\|_F^2 \leq P_T.
\end{aligned}
\end{equation}
We present two precoding schemes tailored for problem \eqref{problem:precoding:elmmse}: Data-dependent precoding (DDP) scheme and data-independent precoding (DIP) scheme. 
\subsubsection{Data-Dependent Precoding}
Let $\left\{\bm{S}_m\right\}_{m=1}^M$ denote a set of $M$ i.i.d. Gaussian data realizations. In the monostatic mode, each $\bm{S}_m$ is known to both the ISAC Tx and the sensing Rx. As a result, the precoding matrix $\bm{W}$ in problem \eqref{problem:precoding:elmmse} can be designed as a function of $\mathbf{S}$ across all data realizations, which we denote as $\bm{W}_m:=\bm{W}(\bm{S}_m)$. The corresponding optimization problem then takes the following form: 
\begin{equation}\label{LMMSE_SubProblem}
\begin{aligned}
\min_{\bm{W}_m}~&  \mathrm{Tr}\left\{\Big(\bm{R}_H^{-1} +\frac{1}{\sigma_s^2 N_s} \bm{W}_m\bm{S}_m\bm{S}_m^H\bm{W}_m^H\Big)^{-1}\right\}\\
\text{\normalfont s.t. }~&\|\bm{W}_m\|_F^2 \leq P_T.
\end{aligned}\end{equation}

%
%

For each given data realization $\bm{S}_m$, problem \eqref{LMMSE_SubProblem} admits a closed-form solution, as shown in \cite[Theorem 1]{10596930}. Specifically, let $\bm{S}_m = \bm{U}_{m} \bm{\varSigma}_{m}  \bm{V}_{m}^H$ denote the singular value decomposition (SVD) of $\bm{S}_m$ and define the matrix $\bm{\varPi}_0$ as 
\begin{equation}\label{permutation_mat}
 \bm{\varPi}_0=\left[\begin{array}{cccc}
 0 & 0 & \cdots & 1 \\
 0 & \cdots & 1 & 0 \\
 \vdots & \vdots & \vdots & \vdots \\
 1 & 0 & \cdots & 0
 \end{array}\right].
 \end{equation} 
Then, the modified water-filling solution of problem \eqref{LMMSE_SubProblem} is expressed as \cite{tang2011waveform}:
\begin{align}\label{Opt_DDP}
\bm{W}_{m}^{\mathsf{opt}} = \bm{Q}\big[ ( \mu_n \bm{\varTheta}_m^{\frac{1}{2}} - \bm{B}_m )^{+} \big]^{\frac{1}{2}} \bm{\varPi}_0\bm{U}_{m}^H,~ m = 1,2, \dots, M,
\end{align}
where $\bm{\varTheta}_m = \frac{1}{\sigma_s^2 N_s} \bm{\varPi}_0\bm{\varSigma}_{m}\bm{\varSigma}_{m}^{T}\bm{\varPi}_0$, $\bm{B}_m = (\bm{\varLambda}\bm{\varTheta}_m )^{-1}$, and $\mu_m$ is a parameter chosen to satisfy the transmit power constraint $\| \bm{W}_{m}^{\mathsf{opt}} \|_F^2 = P_T$.

It is evident from the expression of $\bm{W}_{m}^{\mathsf{opt}}$ in \eqref{Opt_DDP} that the precoding matrix $\bm{W}_m$ depends on the data realization $\bm{S}_m$. Therefore, this scheme is referred to as the data-dependent precoding (DDP). Since the DDP scheme is designed adaptively based on the instantaneous data realization, it generally achieves the minimum ELMMSE; however, this comes at the cost of high computational complexity.    
\subsubsection{Data-Independent Precoding}
Different from the DDP scheme, the DIP scheme aims to find a precoder $\bm{W}$ that is independent of the signal realization. Given the fact that the closed-form expression of ${f}_{\mathsf{ELMMSE}}$ is non-obtainable, the data-independent precoder can be obtained by applying the stochastic gradient descent (SGD) algorithm to solve problem \eqref{problem:precoding:elmmse} offline, providing a favorable tradeoff between estimation performance and computational complexity. Below we present the SGD algorithm in detail.

Let ${f}(\bm{W} ; \mathbf{S})$ be defined as in \eqref{Cond_MMSE}. The gradient of ${f}(\bm{W};\mathbf{S})$ with respect to the variable $\bm{W}$ at a given point $\bm{W}_0$ is
\begin{align}\label{SGD_Gradient}
\nabla{f}(\bm{W}_0;\mathbf{S}) =  \frac{-\left(\bm{R}_H^{-1} + \frac{1}{\sigma_s^2 N_s}\bm{W}_0\mathbf{S}\mathbf{S}^H\bm{W}_0^H\right)^{-2}\bm{W}_0 \mathbf{S}\mathbf{S}^H}{\sigma_s^2 N_s}. 
\end{align}     
Accordingly, the gradient of ${f}_{\mathsf{ELMMSE}}$ with respect to $\bm{W}$ at point $\bm{W}_0$ is given by
\begin{equation}\label{gradientJ}
\begin{aligned}
\nabla{f}_{\mathsf{ELMMSE}}(\bm{W}_0)&=\mathbb{E}\left\{\nabla{f}(\bm{W}_0;\mathbf{S})\right\},
\end{aligned}
\end{equation} which again has no closed-form expression. Towards that end, the key idea behind SGD is to approximate the true gradient in \eqref{gradientJ} by the gradient evaluated at a mini-batch of samples, given by
\begin{align}\label{min_batch gradient}
\hat{\nabla} {f}_{\mathsf{ELMMSE}}(\bm{W}_0) = \frac{1}{|\mathcal{D}|} \sum_{\bm{S} \in \mathcal{D}} {\nabla} {f}(\bm{W}_0; \bm{S}),
\end{align} where $\mathcal{D}$ denotes number of Gaussian samples generated at point $\bm{W}_0.$ 
At the $r$-th iteration of the projected SGD algorithm, the precoding matrix $\bm{W}$ is updated as
\begin{align}\label{BF_SGDupdate}
\bm{W}^{(r+1)} = \mathsf{Proj}\left\{\bm{W}^{(r)} - \varepsilon^{(r)} \hat{\nabla} {f}_{\mathsf{ELMMSE}}(\bm{W}^{(r)})\right\},
\end{align}     
where $\varepsilon^{(r)}$ is the stepsize (or ``learning rate'') and $\mathsf{Proj}\{\cdot\}$ is the projection operator onto the feasible set of problem \eqref{problem:precoding:elmmse}, i.e., the ball constraint.

Some remarks regarding the above SGD algorithm are in order. First, to ensure the convergence of the (projected) SGD algorithm, the nonnegative stepsizes $\varepsilon^{(r)}$ in \eqref{BF_SGDupdate} must be chosen to satisfy the following conditions \cite{eon1998online,liuan2019stochastic}:
\begin{equation}
    \sum_{r=1}^{\infty}\varepsilon^{(r)} = \infty~\text{and}~\sum_{r=1}^{\infty}\left|\varepsilon^{(r)}\right|^2 < \infty.
\end{equation}
Second, increasing the mini-batch size $|\mathcal{D}|$ can reduce the variance of the error in approximating the true gradient, thereby improving numerical performance in terms of computational efficiency and robustness. However, this comes at the cost of having to compute a larger number of local gradients. By the law of large numbers, as the mini-batch size tends to infinity, the gradient in \eqref{min_batch gradient} converges to the true gradient in \eqref{gradientJ}, and the (projected) SGD algorithm effectively reduces to the (projected) GD algorithm. Finally, incorporating the moment information into the SGD algorithm can accelerate its convergence. In particular, a modified version of SGD with momentum was introduced in \cite{10596930} to solve problem \eqref{problem:precoding:elmmse} in the presence of complex unknown variables.

\begin{figure}[!t]
	\centering
	\includegraphics[width = \columnwidth]{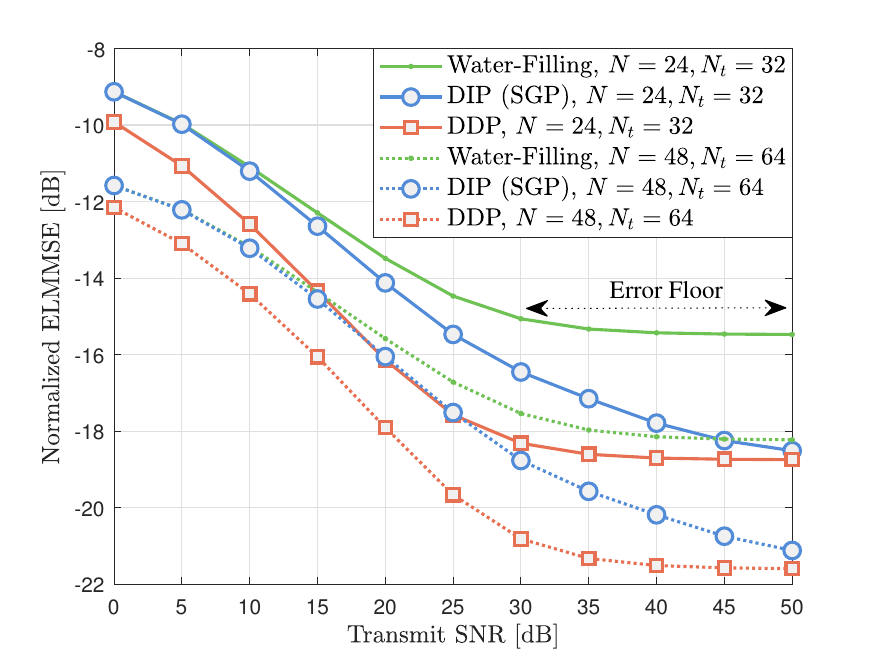}
	\caption{Achievable sensing performance of the DDP and DIP schemes compared with the conventional water-filling approach under Gaussian signaling, where the $N_t = 32, N = 24$ and $N_t = 64, N = 48$ cases are illustrated.}
    \label{Sensing_only_precoding}
\end{figure}

\subsubsection{Example}
We provide an example in Fig. \ref{Sensing_only_precoding} to show the superiority of MIMO precoding designs dedicated to random signals under sensing-only scenarios, where we consider two parameter settings with $N_t = N_s = 32, N = 24$ and $N_t = N_s = 64, N = 48$, respectively. The attainable ELMMSE with varying SNR is shown for both DDP and DIP approaches, with the water-filling precoder \eqref{LMMSE_Determinstic_Opt_W} serving as the baseline scheme. The results for all the three methods are averaged over $1,000$ random realizations of i.i.d. Gaussian signal samples. As predicted by the theoretical analysis, the DDP scheme achieves the lowest ELMMSE in general, followed by its DIP counterpart. On the other hand, the water-filling precoder tailored for deterministic training signals suffers from a 3 dB performance loss as well as a severe error floor compared to the DDP and DIP designs, confirming the necessity of taking the data randomness into account.

\subsection{ISAC Precoding Design}\label{subsection_precoding_isac}
In this subsection, we extend the precoding design from the sensing-only scenario discussed in the previous subsection to the ISAC scenario. Throughout this subsection, we assume that the channel matrix $\bm{H}_{c}$ in \eqref{linear_comm_model} is perfectly known. Then the achievable communication rate (in bps/Hz) of the P2P Gaussian channel in \eqref{linear_comm_model} is \cite{goldsmith2005wireless}:
\begin{align}\label{GaussiaN_sate}
R(\bm{W})  = \log \det \left(  \bm{I}_{N_c} + \sigma_c^{-2} \bm{H}_{c} \bm{W} \bm{W}^{H}\bm{H}_{c}^{H} \right).
\end{align}
The precoding design problem in the ISAC system is formulated as the minimization of ELMMSE, subject to communication performance and power budget constraints, as follows:
\begin{equation}\label{problem:precoding:elmmse:isac}
\begin{aligned}
\min_{\bm{W}}~& {f}_{\mathsf{ELMMSE}}(\bm{W})\\
\text{\normalfont s.t. }~& R(\bm{W})\geq R_0,~\|\bm{W}\|_F^2 \leq P_T,
\end{aligned}
\end{equation}  
where $R_0$ corresponds to the communication rate requirement. Again, we present two precoding schemes, namely, DDP and DIP, tailored for problem \eqref{problem:precoding:elmmse:isac}.

\subsubsection{Data-Dependent Precoding}
We follow the same approach and notation as in Sec. \ref{subsection_precoding_sensing}. By introducing an auxiliary variable $\bm{\varOmega }_{m} = \bm{W}_{m}\bm{W}_{m}^H,$ we can rewrite the rate constraint in problem \eqref{problem:precoding:elmmse:isac} as 
\begin{equation}
 \tilde{R}(\bm{\varOmega }_{m}) := \log \det \left(  \bm{I}_{N_c} + \sigma_c^{-2} \bm{H}_{c}\bm{\varOmega }_{m}\bm{H}_{c}^{H} \right) \ge R_0.   
\end{equation}
The data-dependent precoder $\bm{W}$ can then be obtained by solving the following optimization problem with respect to $\bm{W}_n$ for a given data realization $\bm{S}_m:$ \begin{equation}\label{LMMSE_SubProblem_isac}
\begin{aligned}
\min_{\bm{W}_m, \bm{\varOmega }_{m}}~&  {f}(\bm{W}_m ; \bm{S}_m)\\
\text{\normalfont s.t. }~~\,&\tilde{R}(\bm{\varOmega }_{m})\geq R_0,~\bm{\varOmega }_{m} = \bm{W}_{m}\bm{W}_{m}^H,~\|\bm{W}\|_F^2 \leq P_T.
\end{aligned}\end{equation}

A penalty-based alternating optimization (AO) algorithm has been proposed in \cite{10596930} for solving problem \eqref{LMMSE_SubProblem_isac}. The algorithm essentially applies the AO algorithm to solve the penalized version of problem  \eqref{LMMSE_SubProblem_isac} as follows:
\begin{equation}\label{LMMSE_SubProblem_penalty_isac}
\begin{aligned}
\min_{\bm{W}_m, \bm{\varOmega }_{m}}~&  {f}(\bm{W}_m ; \bm{S}_m)+\frac{\rho}{2}\|\bm{\varOmega }_{m}-\bm{W}_{m}\bm{W}_{m}^H\|_F^2\\
\text{\normalfont s.t. }~~\,&\tilde{R}(\bm{\varOmega }_{n})\geq R_0,~\|\bm{W}\|_F^2 \leq P_T,
\end{aligned}\end{equation} where $\rho>0$ is the penalty parameter. Specifically, problem \eqref{LMMSE_SubProblem_penalty_isac} with respect to the variable $\bm{\varOmega }_{m}$ is convex and can be solved efficiently; the corresponding subproblem with respect to the other variable $\bm{W}_{m}$ can be addressed using the projected gradient descent algorithm. For more details on the proposed penalty-based AO algorithm, please refer to \cite[Sec. IV]{10596930}. 

\begin{figure}[!t]
	\centering
	\includegraphics[width = \columnwidth]{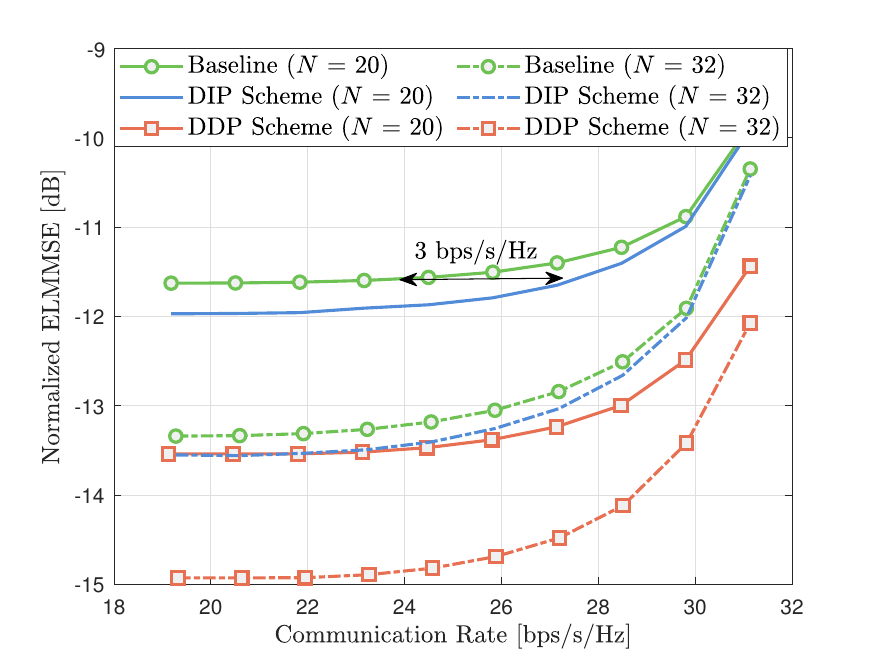}
	\caption{The S\&C performance tradeoff under different precoding designs, with the parameter setting $N_t = N_s = 32, N = 20\;\text{and}\;32$, and $\text{SNR} = 15\;\text{dB}$.}
    \label{ISAC_precoding}
\end{figure}

\subsubsection{Data-Independent Precoding}
By introducing the auxiliary variable $\bm{\varOmega} = \bm{W}\bm{W}^H$ and adding a penalty term for this equality constraint to the objective function, we obtain the following penalized version of problem \eqref{problem:precoding:elmmse:isac}: 
\begin{equation}\label{problem:precoding:elmmse:isac:dip}
\begin{aligned}
\min_{\bm{W},\bm{\varOmega }}~& \mathbb{E}_{\textbf{S}} \left\{{f}(\bm{W};\textbf{S})\right\}+\frac{\rho}{2}\|\bm{\varOmega }-\bm{W}\bm{W}^H\|_F^2\\
\text{\normalfont s.t. }~& \tilde{R}(\bm{\varOmega})\geq R_0,~\|\bm{W}\|_F^2 \leq P_T.
\end{aligned}
\end{equation} This problem can be solved in a similar manner to problem \eqref{LMMSE_SubProblem_penalty_isac} using the AO algorithm. The only difference is that the projected SGD algorithm is employed to solve the subproblem with respect to the variable $\bm{W}.$  
\subsubsection{Example}
We illustrate the tradeoff between S\&C performance of DDP and DIP schemes in Fig. \ref{ISAC_precoding}, under $N_t = N_s = 32, N = 20\;\text{and}\;32$, and $\text{SNR} = 15\;\text{dB}$. The baseline technique here is the precoding design that minimizes the deterministic LMMSE \eqref{deterministic_LMMSE} subject to the communication rate and power constraints. It can be clearly observed that the DIP design acquires 3bps/Hz communication rate improvements over the baseline method for $N = 20$, while achieving the same ELMMSE for sensing. Moreover, DDP achieves more than $1.5$ dB reduction in the ELMMSE compared to the DIP, while satisfying the required communication rate.

\section{Open Problems and Future Directions}\label{open_problems}
In this section, we highlight the open problems in sensing with random communication signals, and identify promising directions for future research in this area.
\subsection{Open Problems}
\subsubsection{2D Ambiguity Function Characterization}
Current research primarily utilizes the ACF to analyze the sensing performance of communication-centric ISAC signals. However, this only provides a partial view of the system's performance. Specifically, the ACF represents the \textit{zero-Doppler slice} of the ambiguity function (AF) \cite{levanon2004radar}, which only captures the multi-target ranging performance under static or quasi-static conditions. For future 6G networks, which are expected to support both S\&C services for numerous moving targets and users, the 2D AF, encompassing both delay and Doppler parameters, will be a more relevant and proper metric for evaluating the sensing performance of random ISAC signals. However, given that the 2D AF is derived from the 2D auto-correlation of the random ISAC signal in the time-frequency domain, analyzing its statistical properties under arbitrary signaling formats remains a significant challenge.

\subsubsection{Mismatched Filtering and Sparse Recovery under Random ISAC Signaling}
In this tutorial, as well as in most of the existing literature, sensing signal processing is performed using the MF framework. In this approach, the received echo signal is convolved with a replica of the transmitted ISAC signal to form a range profile, and target detection is based on localizing the resulting peaks. While MF is known to maximize the target’s SNR at each peak, it may not be optimal for reducing sidelobe levels caused by the data payload. To address this, mismatched filtering (MMF) could be explored as a more general method to further enhance the sensing performance, which conceives the impulse response of the filter as a nonlinear function of the transmitted ISAC signal. One example of the MMF is the \textit{Reciprocal Filtering} \cite{Keskin2024fundamental}, which performs element-wise division of the echo signal in the frequency domain. While the Reciprocal Filter can effectively eliminate sidelobes generated by random data, resulting in a clean ``iceberg'' without any ``sea level'', it may suffer from the SNR reduction by amplifying the noise. Thus, it is essential to develop novel MMF techniques that balance sidelobe suppression and SNR loss under random ISAC signaling. Furthermore, sparse recovery techniques, such as matching pursuit algorithms and their variants, can be employed in this context to enhance sensing resolution by exploiting the inherent sparsity of radar targets \cite{eldar2012compressed}. However, a comprehensive investigation is required to address the challenges posed by the randomness of ISAC signals and its impact on the performance of these algorithms.

\subsubsection{Adaptive Modulation}
In Sec. \ref{PCS_sec} it was shown that sensing favors signals with reduced power variability or even constant modulus signals, at the expense of communication rates. These signals however tend to have higher power efficiency and offer higher SNRs, as well as the opportunity to exploit wireless interference \cite{9035662}. This avails the potential to recover some of the rate loss through adaptive modulation (AM) schemes \cite{950343,7295626} and constructive interference (CI) exploitation \cite{7103338}. On one hand, this offers the opportunity to shift the S\&C tradeoffs from the ones showed above to more favorable communication performance, and in this way better secure the communication QoS in the communication-centric ISAC scenarios. On the other hand, such approaches would necessitate the development of new AM approaches, co-designed with the ISAC signaling overviewed in this paper, and ISAC-tailored CI approaches founded on the constellation shaping above.

\subsubsection{Sensing with Channel-Coded Signals}
Channel coding is a critical component of modern communication systems, which adds redundancy to information bits to reduce or correct decoding errors. Most current studies on communication-centric ISAC systems focus on uncoded signals, where constellation symbols are i.i.d. drawn from predefined codebooks. While the effects of modulation schemes, constellation designs, pulse shaping filters, and MIMO precoders on the sensing performance have been investigated in this tutorial, the impact of channel coding remains largely unexplored \cite{9834554}. Therefore, it would be valuable to evaluate the sensing performance of random ISAC signals under various practical channel coding schemes, such as Turbo, LDPC, and Polar codes, and to optimize these codes for achieving a balanced S\&C performance.

\subsection{Future Directions}
\subsubsection{Networked Sensing and ISAC with Communication Signals} 
On top of the P2P ISAC setting considered in this paper, networked sensing and ISAC represent a transformative approach to enabling seamless sensing and communication across large areas \cite{10735119,10726912,10769538,10380513,9842350}. By leveraging collaboration among BSs and distributed mobile devices, networked ISAC is particularly promising for applications like UAV-enabled low-altitude economy and smart transportation systems \cite{9916163,Xujie2024LAE}. Unlike their conventional P2P monostatic and bistatic counterparts, networked sensing and ISAC encounter unique challenges, particularly in interference management. In particular, the simultaneous transmissions from distributed nodes can significantly degrade both S\&C performance across the network, and the inherent randomness of communication signals further complicates the analysis and management of interference. To tackle this challenge, comprehensive research is needed on BSs' synchronization, adaptive BS clustering and scheduling, collaborative precoding, and network-level joint resource allocation. For instance, BSs can be dynamically grouped into clusters, where joint precoding within each cluster can leverage cross-link interference as beneficial signals, while interference coordination cross clusters is crucial to mitigate inter-cluster interference. Based on the availability of data information, multiple BSs can employ the DDP and DIP techniques to improve the sensing performance without compromising communication quality. In such cases, the joint optimization of data-dependent and data-independent precoders across multiple BSs becomes crucial, with distributed algorithm design playing a key role in achieving enhanced performance while minimizing signaling overheads.

\subsubsection{Secure ISAC with Communication Signals}
The move to pervasive sensing through the ISAC infrastructure opens the door to entirely new security vulnerabilities over the wireless network \cite{9755276}: i) {\textit{Data-Security}}: the inclusion of data into the probing ISAC signal makes it prone to eavesdropping from potentially malicious radar targets, and with high signal powers typically used for target illumination. Even if the data itself is encrypted, simply detecting the existence of a communication link can jeopardize communication privacy \cite{9737364,9199556}. ii) {\textit{Sensing-Privacy}}: The sensing functionality introduced by the wireless network can be adversely exploited by malicious nodes to independently sense potentially sensitive information about the environment \cite{10587082}. This is an entirely new vulnerability that one never had to worry about in a cellular network. As there is no data link - this is the ability of a malicious node to independently sense its environment - higher layer security approaches are inapplicable. The severity of threat necessitates a new generation of PHY security solutions tailored for ISAC. The constellation, pulse shaping, and precoding design overviewed in this paper needs to be tuned for PHY security and can play a key role in protecting against both data eavesdropping and adversary sensing. Their co-design with classical PHY security approaches such as artificial noise design, jamming, cooperative security remains virtually unexplored. Most importantly, while data security in ISAC is being explored theoretically \cite{10122612}, the realm of sensing privacy lacks an information theoretic framework with which to design metrics, signal processing solutions and transmission mechanisms.

\subsubsection{Sensing and ISAC with Artificial Intelligence (AI)} While this paper focuses on the information theory and signal processing aspects of sensing and ISAC, AI has recently emerged as a key enabler, particularly in processing sensing and ISAC signals for recognition tasks \cite{CuiEdge2024,zhu2023pushing}. Deep learning algorithms, for instance, are increasingly being utilized for applications such as posture and activity recognition. The integration of AI with sensing and ISAC introduces both new challenges and exciting opportunities. A key challenge arises when random communication signals are employed, as designing AI algorithms capable of effectively processing the resulting echo signals becomes complex. A promising approach to address this is to combine well-established model-driven radar signal processing methods with innovative data-driven AI techniques. Another significant challenge is defining new sensing performance metrics (e.g., recognition accuracy), and understanding their relationship with the design parameters like modulation types and covariance matrices of transmitted communication signals to guide system optimization. The incorporation of AI makes establishing quantitative connections between them especially difficult. Despite these challenges, AI also offers powerful tools to address these issues. AI techniques can model complex, nonlinear relationships between sensing performance metrics and signal parameters, providing insights that are difficult to obtain through traditional methods. Furthermore, AI can optimize a variety of functional blocks, e.g., input distribution of the constellation, through end-to-end learning \cite{Geiger2025AIPCS}, making it an invaluable asset in advancing sensing and ISAC technologies.

\subsubsection{Integrated Sensing, Communication, and Powering (ISCAP) with Communication Signals} In addition to supporting sensing and communication, radio signals can wirelessly deliver energy to power low-power devices such as sensors and IoT devices through wireless power transfer (WPT). With spectrum resources becoming more limited, future wireless networks are anticipated to combine sensing, communication, and WPT, creating multi-functional ISCAP networks \cite{10382465,Xujie2024ISCAP}. ISCAP presents new challenges in designing signal waveforms and optimizing communication signals to balance the tradeoff among sensing, communication, and WPT. Unlike traditional ISAC systems, ISCAP must address the unique requirements of WPT, where the energy harvesting efficiency depends heavily on the characteristics of the transmitted waveform. Due to the non-linear radio frequency (RF)-to-direct current (DC) conversion process in energy harvesting devices, waveforms with high peak-to-average power ratio (PAPR) are typically favored to maximize power transfer efficiency. To overcome these challenges, optimizing modulation, waveform design, and beamforming is crucial. While some initial studies have explored waveform and beamforming designs in simplified ISCAP scenarios, a comprehensive system-level analysis and design remain underdeveloped. For instance, analyzing WPT performance under various waveforms, such as OFDM, CDMA, and OTFS, while accounting for practical energy harvesting constraints, offers a promising research direction. Similarly, extending data-dependent and data-independent precoding techniques for ISCAP is also interesting.

\section{Conclusions}\label{conclusions}
This tutorial paper has examined recent developments in the field of communication-centric ISAC transmission, which maximizes the resource utilization efficiency by leveraging random data payload signals for both S\&C tasks. We first discussed the information-theoretic foundation of ISAC, emphasizing the necessity of developing signal processing techniques tailored for random ISAC signals. Following this, we reviewed the core models and methodologies for communication-centric ISAC systems, with a particular focus on analyzing the statistical properties of the ACF of ISAC signals, which is critical for evaluating multi-target sensing performance. As a step further, a significant part of the discussion was dedicated to the design principles for key components of ISAC systems, including modulation schemes, constellation design, and pulse shaping filters. Here, we highlighted the importance of optimizing the sensing functionality without sacrificing the communication performance, or in some cases, developing a scalable tradeoff that supports both functionalities. On top of that, we also explored the advancements in MIMO systems, particularly in the context of dedicated sensing and ISAC precoding techniques conceived for random data payload signals. Finally, the paper concluded by identifying several open research challenges and outlining future directions in communication-centric ISAC transmission. It is our hope that this work will help guide the ongoing development of ISAC air interface technologies that are compatible with the current cellular networks, and contribute to the standardization and implementation of ISAC in future 6G wireless networks.

\bibliographystyle{IEEEtran}
\bibliography{IEEEabrv,references_SPM,references,database}

\end{document}